\pdfoutput=1 
\documentclass{JINST}

\usepackage{subfigure}
\usepackage{multirow}
\usepackage{epsfig}

\usepackage{lineno}

\title{Krypton and radon background in the PandaX-I dark matter experiment}

\author{
Shaoli Li$^a$,
Xun Chen$^a$,
Xiangyi Cui$^a$,
Changbo Fu$^a$,
Xiangdong Ji$^{a,b,c,d}$,
Qing Lin$^a$\thanks{Now at Department of Physics, Columbia University},
Jianglai Liu$^a$\thanks{Corresponding author, jianglai.liu@sjtu.edu.cn.},
Xiang Liu$^{a,b}$,
Andi Tan$^c$,
Xuming Wang$^a$,
Mengjiao Xiao$^{c,b}$,
Pengwei Xie$^{a}$\\
\llap{$^a$}INPAC and Department of Physics and Astronomy, Shanghai Jiao Tong University, \\
Shanghai Key Laboratory for Particle Physics and Cosmology, Shanghai, 200240, China\\
\llap{$^b$}Center of High Energy Physics, Peking University, Beijing, 100080, China\\
\llap{$^c$}Department of Physics, University of Maryland, College Park, MD, 20742, USA\\
\llap{$^d$}T. D. Lee Institute, Shanghai, 200240, China\\
}

\abstract{
We discuss an {\it in-situ} evaluation of the $^{85}$Kr, $^{222}$Rn, and $^{220}$Rn
background in PandaX-I, a 120-kg liquid xenon dark matter direct
detection experiment. Combining with a simulation, their contributions to the 
low energy electron-recoil background in the dark matter search region are obtained. 
}

\keywords{
Krypton; Radon; Background; Dark Matter; Xenon; PandaX
}
 
\begin{document}


\section{Introduction}
Radioactive rare gases, most notably $^{85}$Kr, $^{222}$Rn and $^{220}$Rn, 
are important background in xenon-based dark matter experiments. 
They cannot be removed by conventional hot
getters via chemical reactions and have high mobility to diffuse throughout the 
target volume, which degrade the power of fiducialization. 
The air contains unstable $^{85}$Kr (10.8 year half-life, $\beta$-decay)
whose activity is $\sim$1 Bq/m$^{3}$
produced by the fissions of uranium and plutonium 
in our nuclear age~\cite{KrAbundance}. If air is leaked into the detector, 
$^{85}$Kr would produce $\beta$-decay background in the detector.
Radon gas is produced by the decay of the long-lived $^{238}$U and $^{232}$Th, 
which could be introduced into the detector either externally by an air leak 
or internally by surface emanation from detector materials.
Direct measurement of such background during data taking is critical.

PandaX-I~\cite{pandaxI} was a dark matter search experiment operated at
the China JinPing underground laboratory (CJPL)~\cite{CJPL1, CJPL2, CJPL3} 
using a dual-phase time projection chamber (TPC) with a 120-kg active liquid xenon target.
The apparatus has been described in detail in Ref.~\cite{pandaxI}, so only the key 
aspects relevant to this paper are presented here.
A particle interacting in liquid xenon (LXe) produces prompt scintillation photons (S1).
Some ionized electrons will drift under a drift field defined by the 60 cm diameter 
cathode grid ($-15$~kV) and the gate grid ($-$5~kV) located at the bottom and 
top of the sensitive liquid region, separated by 15~cm. The electrons will
be extracted into the gaseous xenon by a stronger field across the liquid 
surface between the gate grid and the anode mesh (ground), separated by 8~mm, and 
produce proportional electroluminescense photons (S2). The S1 and S2 are 
collected by the photomultiplier (PMT) arrays 
located at the top and bottom of the TPC. When 
in combination with the time separation between S1 and S2, they allow 
three-dimensional reconstruction of the event vertex.
The ratio between the
S2 and S1 for the nuclear recoil (NR) signal produced by dark matter scattering with 
xenon nucleus is significantly lower than that for 
electron recoil (ER) background produced by the 
$\gamma$s or $\beta$s, enabling effective background rejection. $\alpha$ 
events, characteristic for radon decays, can be identified by their discrete 
high value of energy deposition but an even less S2-to-S1 ratio in comparison to the 
NR events. 

The results for dark matter search from 
PandaX-I have been published in Refs.~\cite{pandaxI1st, pandaxI2nd}. 
In this paper, we present
a detailed {\it in-situ} evaluation of the krypton and radon background in the
full exposure (54$\times$80.1 kg-day) dark matter search data in PandaX-I. 
Same as Ref.~\cite{pandaxI2nd}, we have applied the widely used delayed coincidence
to tag these background in the data, as in Refs.~\cite{XENON100_bk,XENON100_Rn,LUX_bk,LUX_Rn,LUX_Rn2}.
In this paper, we updated the corresponding results from Ref.~\cite{pandaxI2nd} due
to the following updates. First, the $\beta$ and $\gamma$ energy cuts for $^{85}$Kr
were changed slightly in comparison to Ref.~\cite{pandaxI2nd}, and in this paper we used
all 80-day data to search for $^{85}$Kr delay-coincidence (Ref.~\cite{pandaxI2nd} only
used the blinded 63-day data). Second, we have updated the signal finding efficiency from
the Monte Carlo (MC) simulation, which has a more realistic treatment for close-by delay-coincidence signals.
Third, we have made accidental background subtraction in all delayed coincidence analysis.
The rest of this paper is organized as follows: Secs.~\ref{sec:Kr}, ~\ref{sec:Rn222}
and ~\ref{sec:Rn220} 
are dedicated to $^{85}$Kr, $^{222}$Rn, and $^{220}$Rn background,
respectively, using characteristic delayed coincidence along their decay chains; 
Sec.~\ref{sec:SingleAlpha} is an alternative radon study based on
single $\alpha$s; and Sec.~\ref{sec:sum} contains the summary.

\section{Krypton background}
\label{sec:Kr}
$^{85}$Kr $\beta$-decays with 99.566\% of the probability
directly into stable $^{85}$Rb with a $Q$-value of 687 keV, and 0.434\% probability
into the 
meta-stable $^{85\rm{m}}$Rb with the maximum $\beta$ energy of 173 keV. 
$^{85\rm{m}}$Rb then
de-excites with a half-life of 1.015 $\mu$s, emitting a $\gamma$ ray of 514~keV. 
The $^{85}$Kr level can therefore be estimated using the $\beta$-$\gamma$ delayed
coincidence signature.

A search on the $\beta$ and $\gamma$ coincidences was performed in each event. 
As described in Ref.~\cite{pandaxIDAQ}, for each PMT channel, the waveform 
data in a 200~$\mu$s window were recorded. To avoid 
ambiguity in the S1 and S2 pairing, we selected events in which 
both the $\beta$ and $\gamma$ made a single scatter. 
For the delayed coincidence events from $^{85}$Kr, about 60\% of them
produced two clean (S1, S2) signals.
The rest contained two S1s but only a single S2 as the time separation between the two
S2 signals could also be less than the width of a typical S2 signal (about 2 $\mu$s).

\begin{figure}[!htbp]
\centering
  \includegraphics[width=0.95\linewidth]{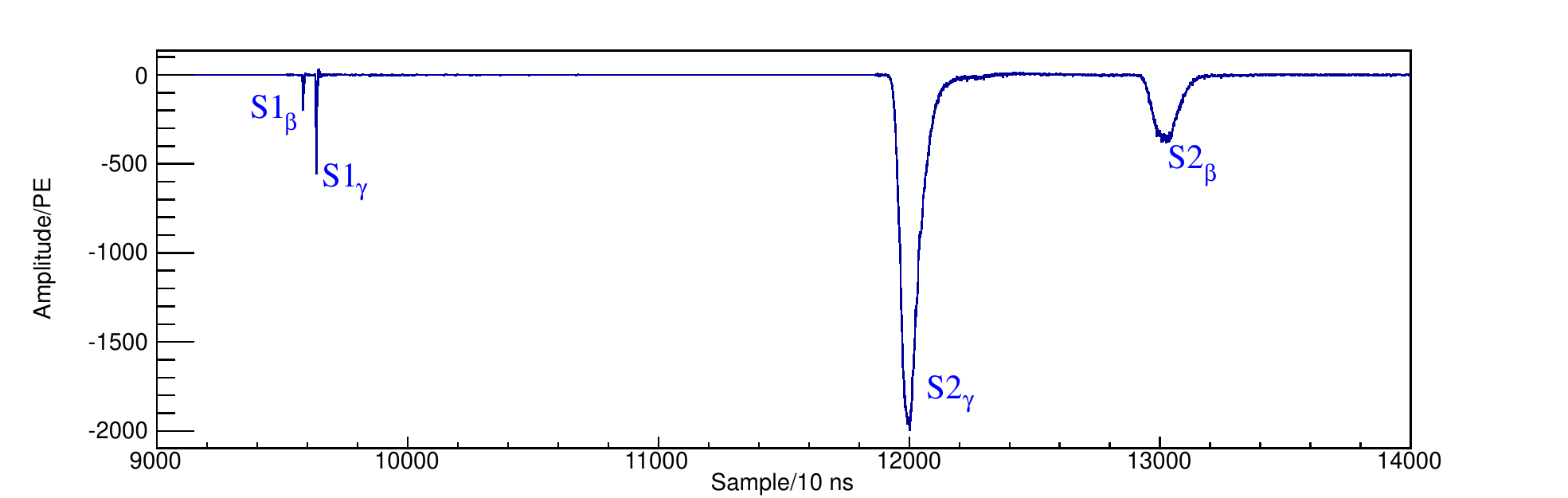}
  \caption{Typical waveform of a $\beta$-$\gamma$ delayed coincidence signal from 
    krypton decay in which two S2s are overlapped.}
  \label{fig:KrWF}
\end{figure}

The selection cuts were applied as follows. The energy values (estimated by S1) of 
the $\beta$ and $\gamma$
were required to be within 20 to 200 keV, and 300 to 700 keV, respectively, and  
the time separation between the two S1s was required to be within 0.3 and 3 $\mu$s. 
An example coincidence waveform is shown in Fig.~\ref{fig:KrWF}.
For events with two S1s and S2s, we also required that both $\beta$ and $\gamma$
were consistent with ER events (within a loose ER cut
obtained from the $\gamma$ source calibration data, cf. Fig.~\ref{fig:Bi214_band_logz}).
Note that to maximize the statistics, all candidate events within the 120 kg sensitive 
target volume were selected.

The cut efficiency was obtained using the PandaX-I MC 
simulation package~\cite{ref:PandaXG4}, 
which was developed based on Geant4~\cite{ref:G4} (v9.6) 
with recommended physics lists for low energy physics and included the
full detector geometry of PandaX-I.  
The delay time cut required that both the $\beta$ and $\gamma$ had 
deposited energy inside the detector, and the time separation between them
was between 0.3 to 3 $\mu$s. The efficiency was estimated to be 56.2\%. 
The energy cut efficiencies on $\beta$ and $\gamma$ were 88.8\% and 72.9\%, 
respectively, therefore the overall cut efficiency was 36.4\%.

\begin{figure}[!htbp]
\centering
\subfigure[Delay time]
{ \label{fig:Kr_candi_deltaT}
  \includegraphics[width=0.45\linewidth]{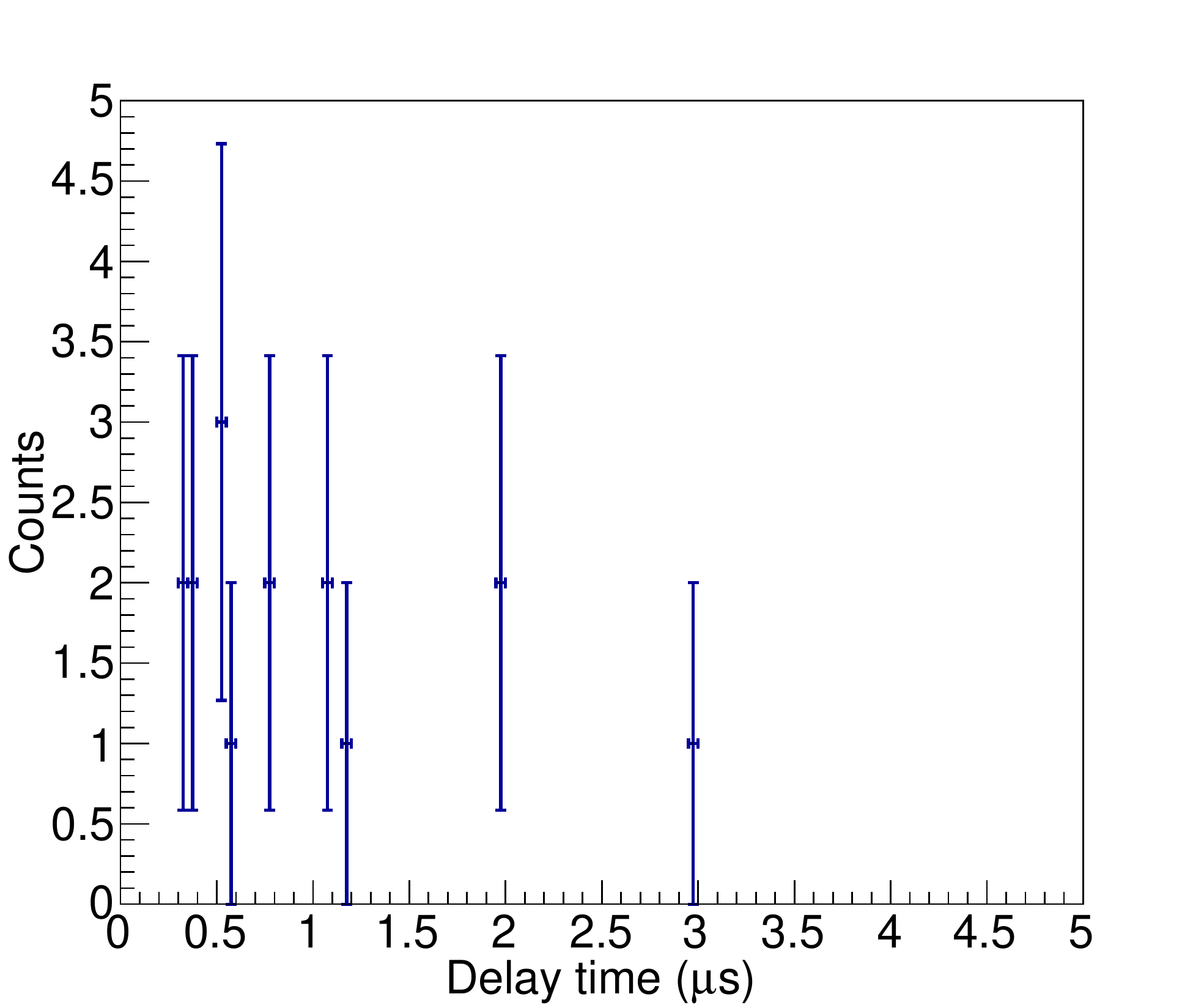}
}
\subfigure[$\Delta L^2$]
{
  \label{fig:Kr_candi_distance}
  \includegraphics[width=0.45\linewidth]{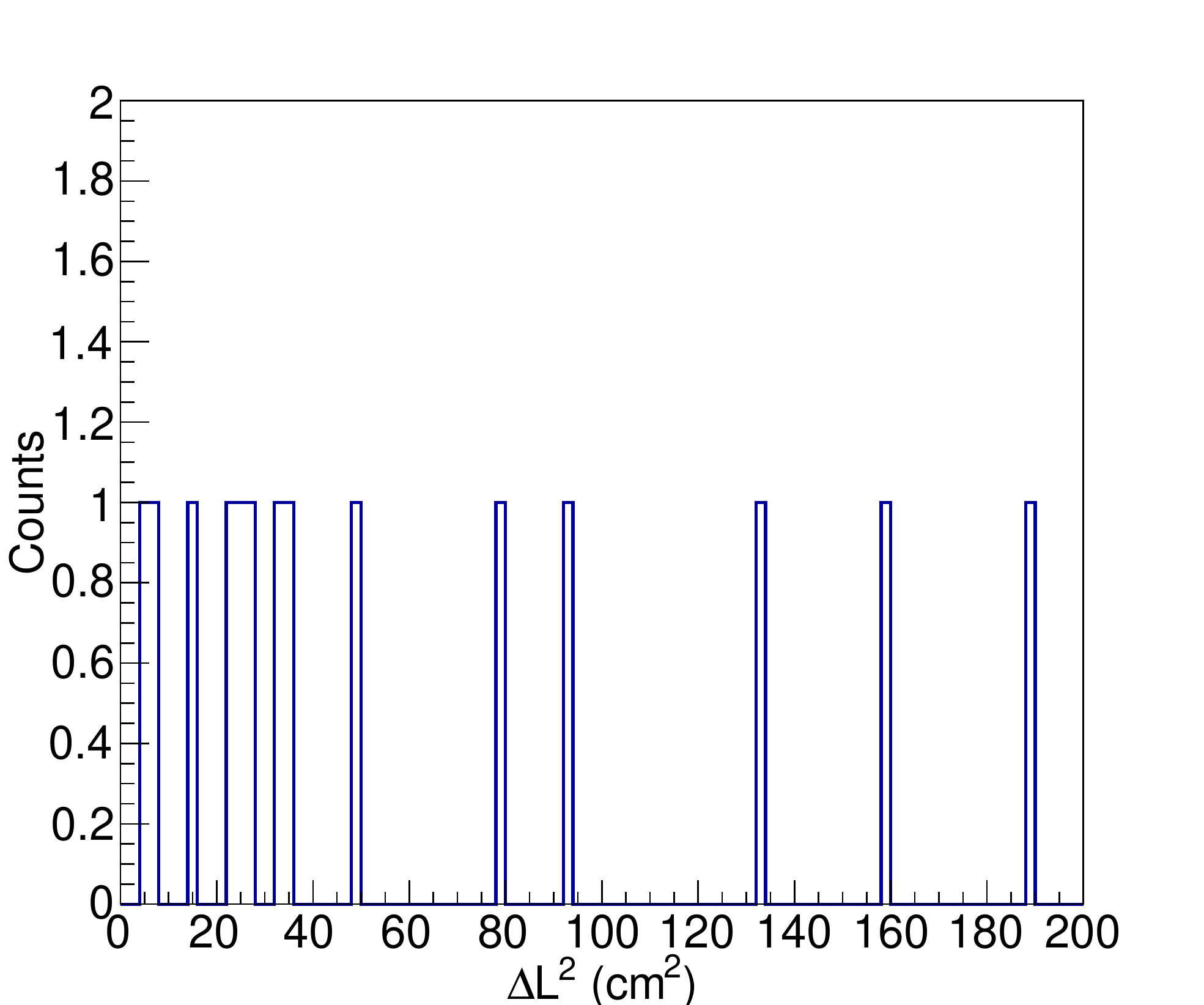}
}
\subfigure[Position]
{
  \label{fig:Kr_candi_position}
  \includegraphics[width=0.45\linewidth]{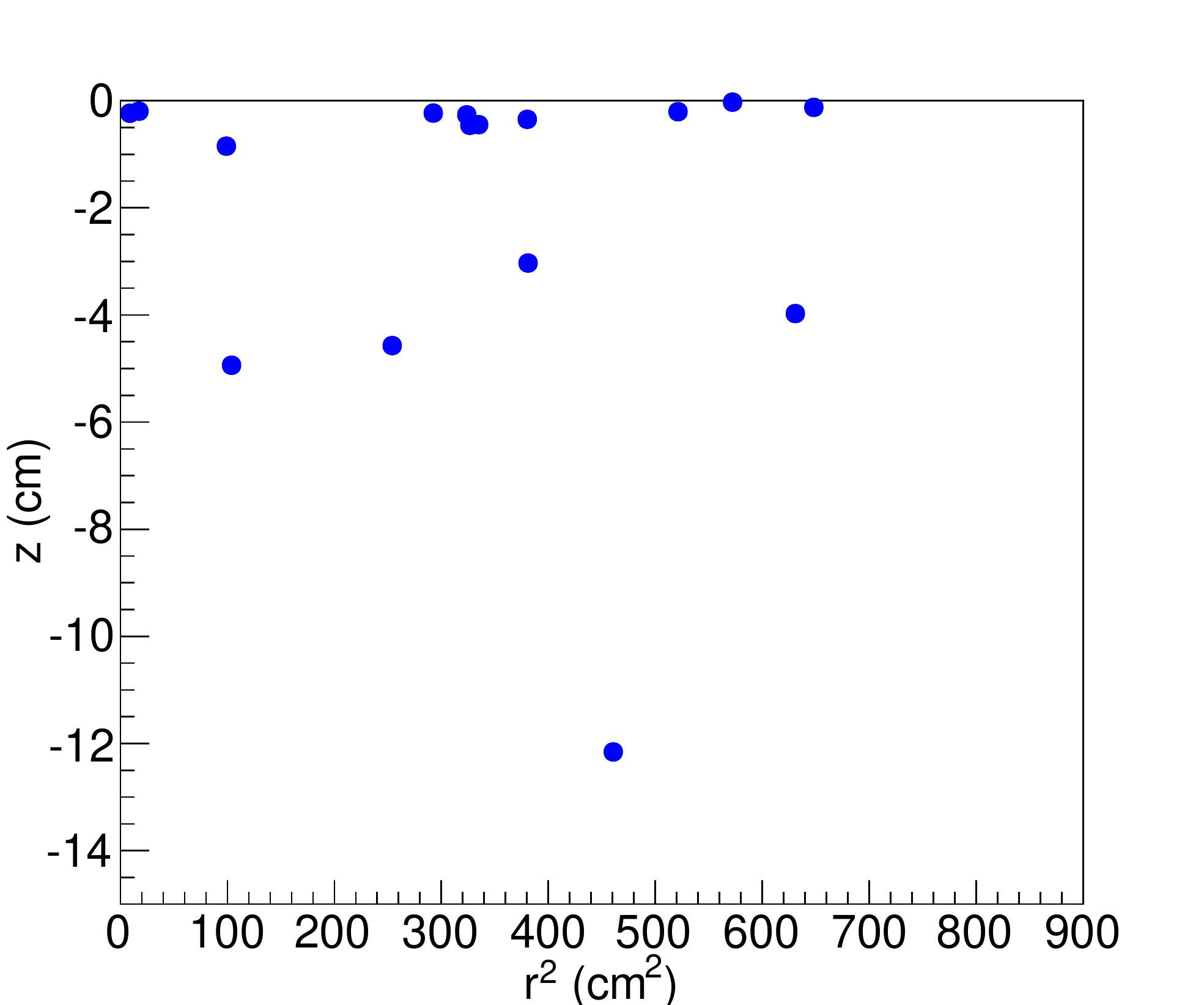}
}
\caption{Distributions in a) Delay time, b) $\Delta L^2$, and c) $z$ vs. $r^{2}$
  for $^{85}$Kr $\beta$-$\gamma$ candidates. Reconstructed position based on
  the S1 signal is used in b) and c) if there is only one S2 signal tagged in the
  delayed coincidence. In c), the $\beta$
  positions are shown.}
\label{fig:Kr_candi_dist}
\end{figure}

After applying the cuts above, 16 events from the data survived 
in the entire 120-kg sensitive volume.
The delay time and the square distance 
($\Delta L^{2}=\Delta x^2+\Delta y^2+\Delta z^2$) between 
the $\beta$ and $\gamma$ are shown
in Figs.~\ref{fig:Kr_candi_deltaT} and \ref{fig:Kr_candi_distance}, both exhibiting 
correlation despite limited statistics. 
The position distribution of the $\beta$s is shown
in Fig.~\ref{fig:Kr_candi_position}.
A number of candidate events happened close to the gas/liquid xenon interface.
This could be due to the fact that the boiling point of krypton ($\sim$120 K)
is lower than that of the xenon ($\sim$165 K).
To be conservative, we assumed that krypton was distributed uniformly in the
detector.
The event selection above contained accidental background arising from
the random coincidence of single $\beta$-like and $\gamma$-like events which
satisfied the selection cuts above.
Based on the dark matter data, we estimated the $\beta$-like and $\gamma$-like event
rates to be 0.24 Hz and 0.4 Hz respectively.
So the random coincidence event rate within the same coincidence window 
(0.3$-$3 $\mu$s) is 2.3$\times$10$^{-2}$ evts/day or 1.8 events in 80.1-day exposure, 
with negligible statistical uncertainty. 

Based on the delayed coincidence candidates
and accidental background, the krypton level in xenon can be estimated by
\begin{equation}
\label{eq:krypton_level}
N_{\rm Kr} =N_{85}/f \,,\quad  N_{85} =\frac{N_{\rm data} - N_{\rm acc} }{\varepsilon_{\rm cut}\cdot BR \cdot (T/\tau)} 
\end{equation}
where $N_{85}$ and $N_{\rm{Kr}}$ are the numbers of $^{85}$Kr and total krypton atoms, 
respectively, related by the $^{85}$Kr abundance $f$ 
(about 2$\times$10$^{-11}$, see Ref.~\cite{KrAbundance}).
$N_{\rm {data}}$ and $N_{\rm {acc}}$ are the number of raw candidates and accidental events
in the data, respectively,
$\epsilon_{\rm{cut}}$ is total cut efficiency, $BR$ is the branching ratio which is 0.434\%,
$T=80.1$-day is the live-time of the dark matter run, and $\tau$ is the 
mean lifetime of $^{85}$Kr (15.52 years).
Using Eq. \ref{eq:krypton_level}, number of $^{85}$Kr atoms is estimated to be
(6.4$\pm$1.8)$\times$10$^5$, leading to a molar 
concentration of 58$\pm$16 part per trillion (ppt) krypton atoms in xenon.

The MC simulation was used to translate the $^{85}$Kr decay rate to 
the background rate observed in the dark matter search region
between 0.5$-$5 keV, leading to a 2.0$\pm$0.6 mDRU
(mDRU = 10$^{-3}$ evt/day/kg/keV) background.
No apparent time dependence was observed in the $\beta$-$\gamma$ coincidence 
candidates, nor did the low energy background in the data exhibit the dependence. 
Therefore, the krypton was likely introduced during
the detector filling period when the detector was underpressurized, 
not by a leak developed during the run.


\section{$^{222}$Rn}
\label{sec:Rn222}
\begin{figure}[h!]
\centering
  \includegraphics[width=.9\textwidth]{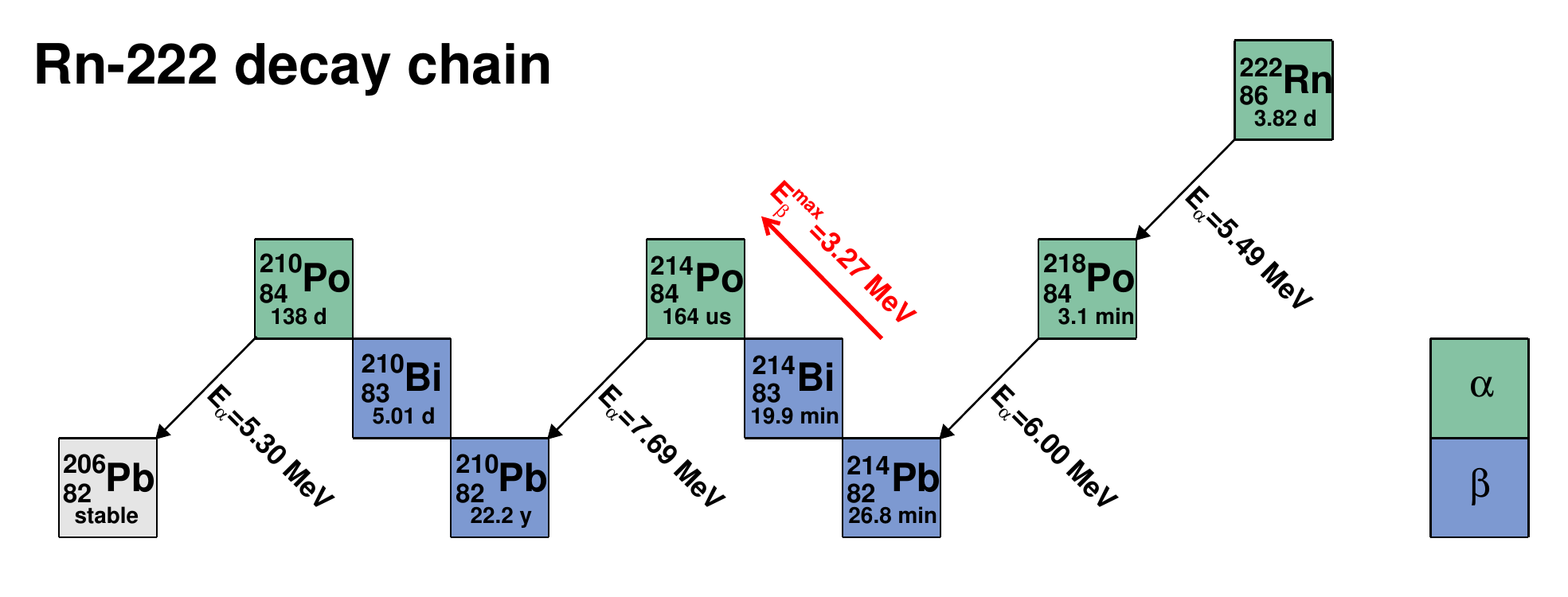}
  \caption{$^{222}$Rn decay chain. The red arrow indicates the $\beta$-$\alpha$ delayed coincidence used 
    in this analysis.}
  \label{fig:Rn222_decay_chain}
\end{figure}

$^{222}$Rn is a decay progeny of $^{238}$U, which is either external airborne
or internal due to surface emanation of radioimpurity in the internal detector components.
$^{222}$Rn has a half-life of 3.82 day, with its decay chain illustrated in
Fig.~\ref{fig:Rn222_decay_chain}.
Along the chain, number of $\beta$-decays
could contribute to the background in the low energy signal region.
Given that the PandaX-I running period was much longer than the 
decay half-life of $^{222}$Rn progenies all the way to $^{210}$Pb,
we have assumed that secular equilibrium was achieved upstream of $^{210}$Pb,
based on which we can estimate the $^{222}$Rn level.

To tag $^{222}$Rn events, we used the $^{214}$Bi-$^{214}$Po $\beta$-$\alpha$ delayed
coincidence events, in which 
$^{214}$Bi emits $\beta$ with a maximum energy of 3.272 MeV,
followed by the $\alpha$ decay of $^{214}$Po with a half-life of 164.3 $\mu$s
and an $\alpha$ energy of 7.69 MeV.

\begin{figure}[!htbp]
\centering
\subfigure[$^{214}$Bi]
{
  \label{fig:Bi214_band_logz}
  \includegraphics[width=.45\linewidth]{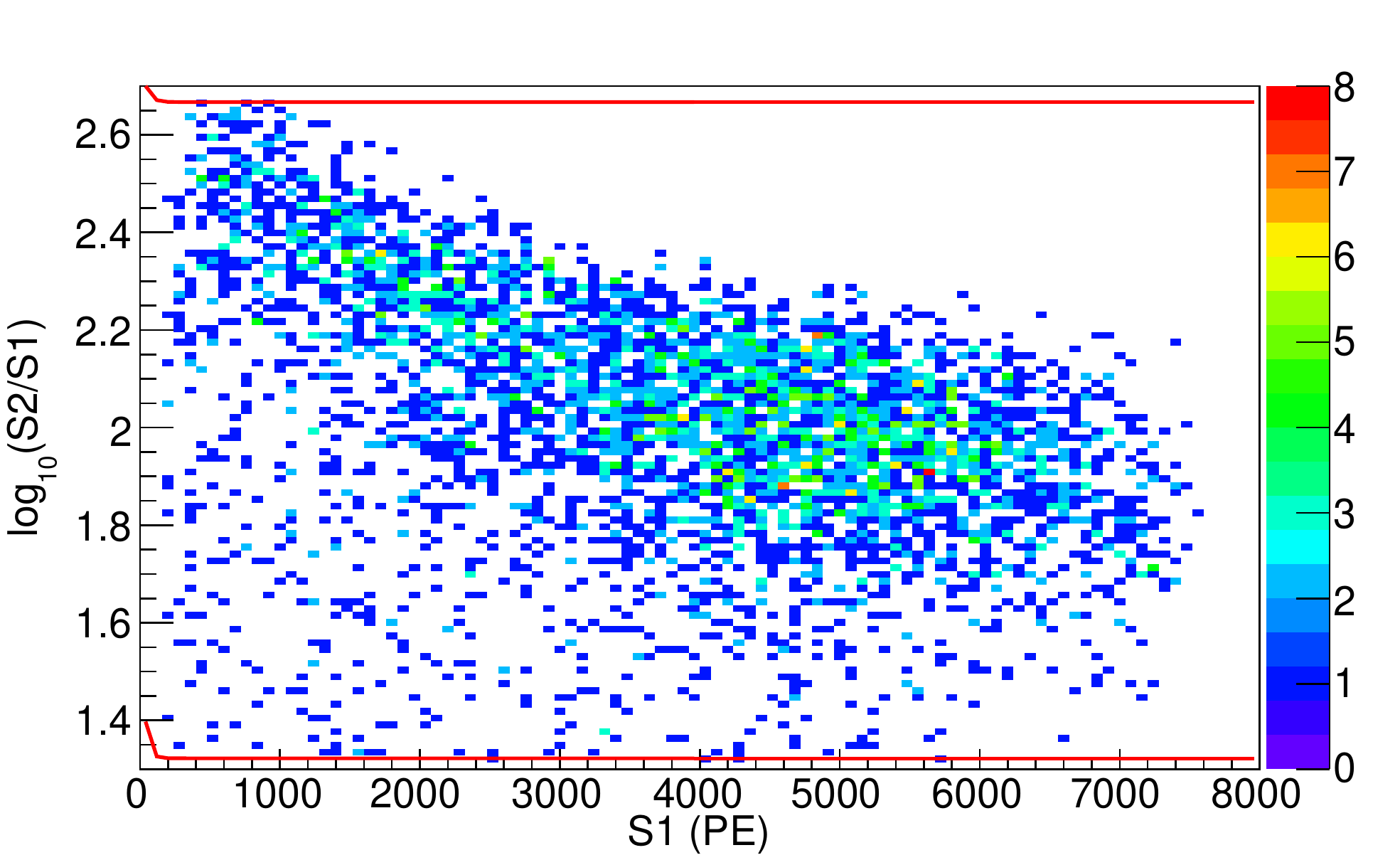}
}
\subfigure[$^{214}$Po]
{
  \includegraphics[width=.45\linewidth]{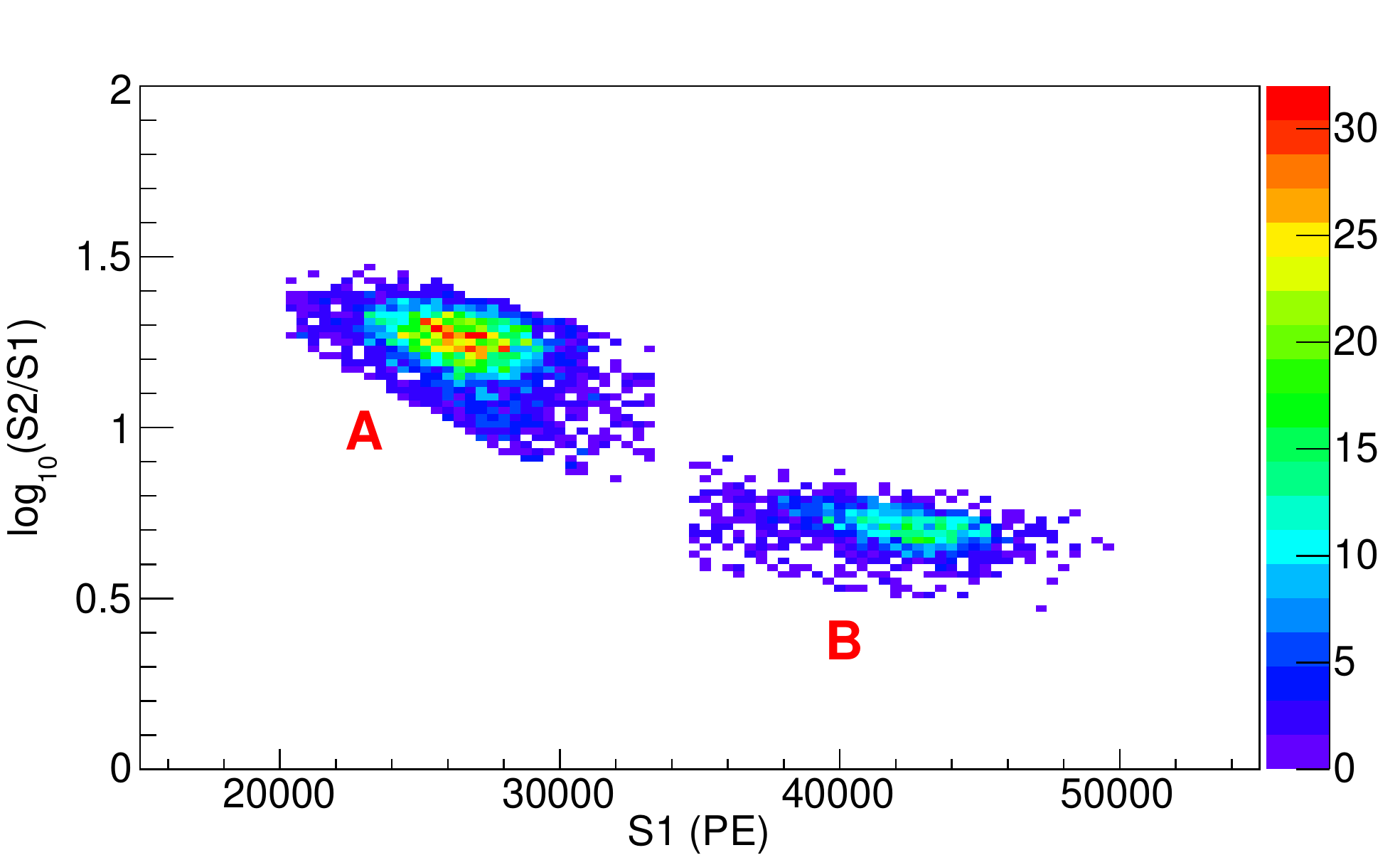}
  \label{fig:Po214_band}
}
\caption{The distribution of log$_{10}$(S2/S1) vs. S1 of $^{214}$Bi (a) and $^{214}$Po (b) of
BiPo2E events. Red curves in (a) are the ER cut defined by the ER calibration.}
\label{fig:Bi214Po214_band}
\end{figure}

As mentioned earlier, the PandaX-I data acquisition window was 
200~$\mu$s \cite{pandaxIDAQ}.
The delayed coincidence signal could either be recorded in one or two adjacent events,
which will be referred to as the ``BiPo1E" and ``BiPo2E" hereafter. 
The $\beta$ energy (scaled from S1) and $\alpha$ energy (see later) 
cuts of 0.1$-$3.5 MeV and 6.0$-$9.5 MeV were applied, respectively.
The time separation cut between the two S1s was set to be 20$-$100 $\mu$s for BiPo1E
or 300$-$1000~$\mu$s for BiPo2E.
In addition, $\beta$ signal should be consistent with an ER signal 
(Fig.~\ref{fig:Bi214_band_logz}).



\begin{figure}[!htbp]
\centering
\subfigure[Delay time]
{
  \includegraphics[width=.45\linewidth]{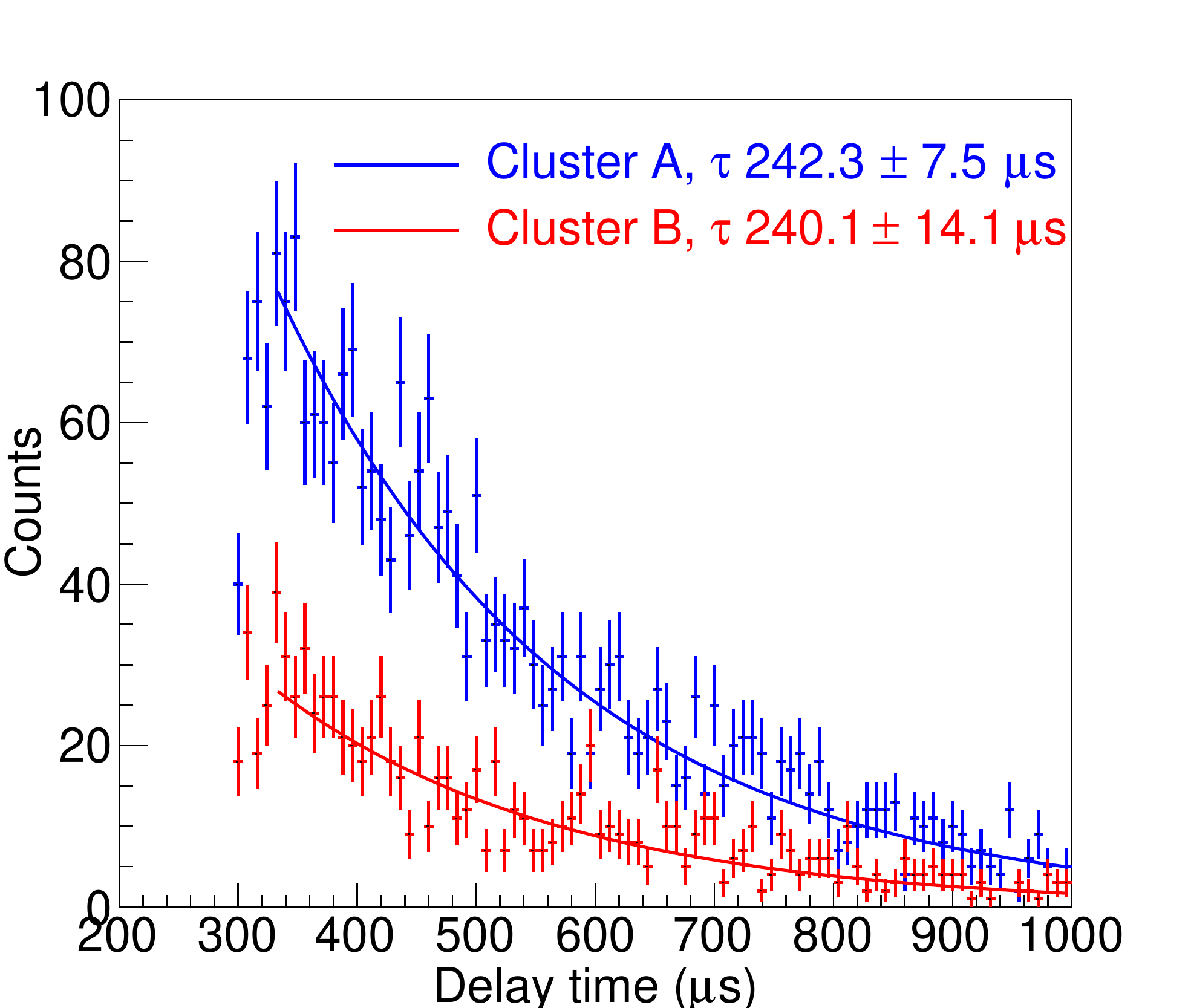}
}
\subfigure[$\Delta L^2$]
{
  \includegraphics[width=.45\linewidth]{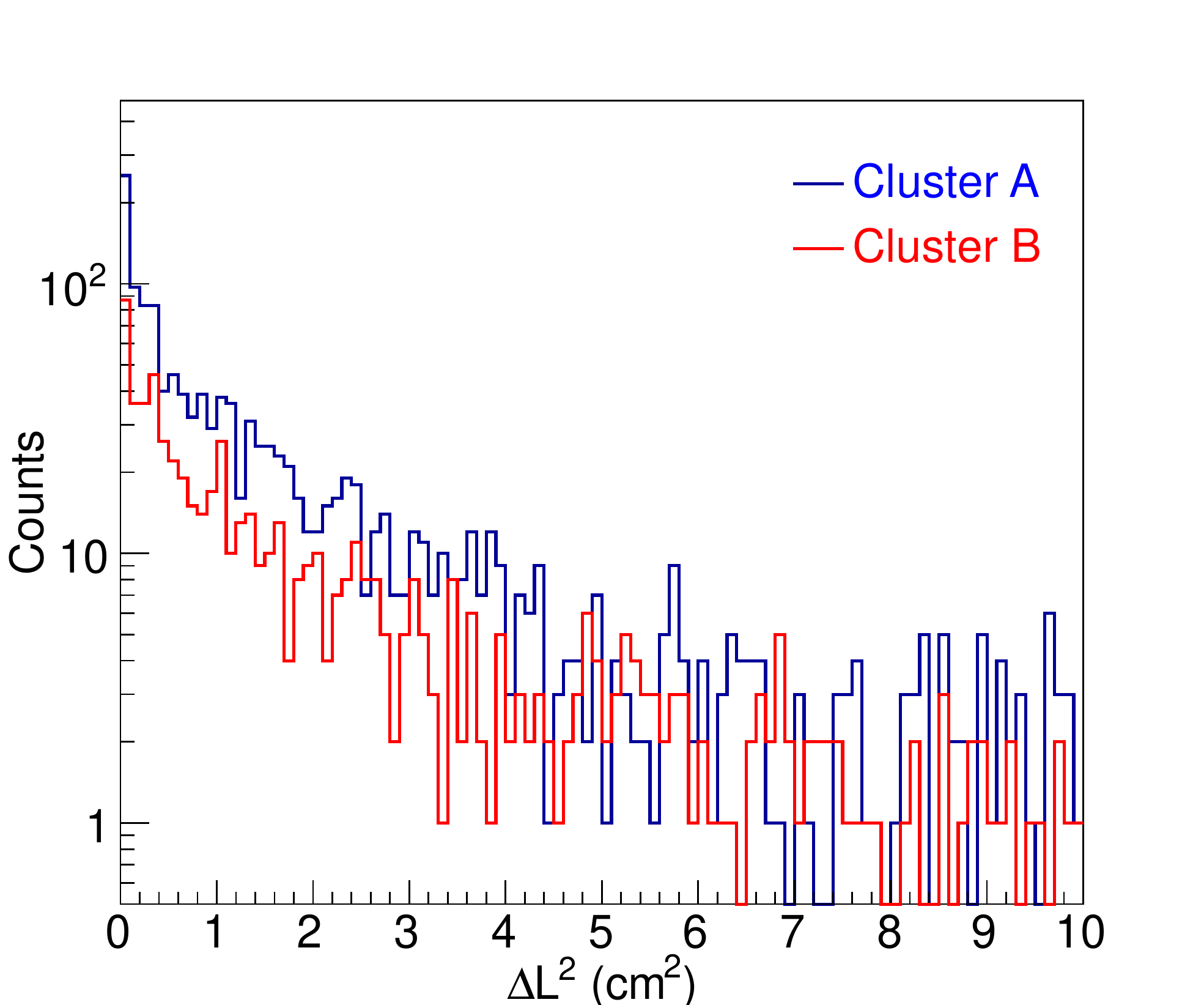}
}
\subfigure[$\alpha$ position]
{
  \label{fig:Rn222_cross_Po214_position}
  \includegraphics[width=.45\linewidth]{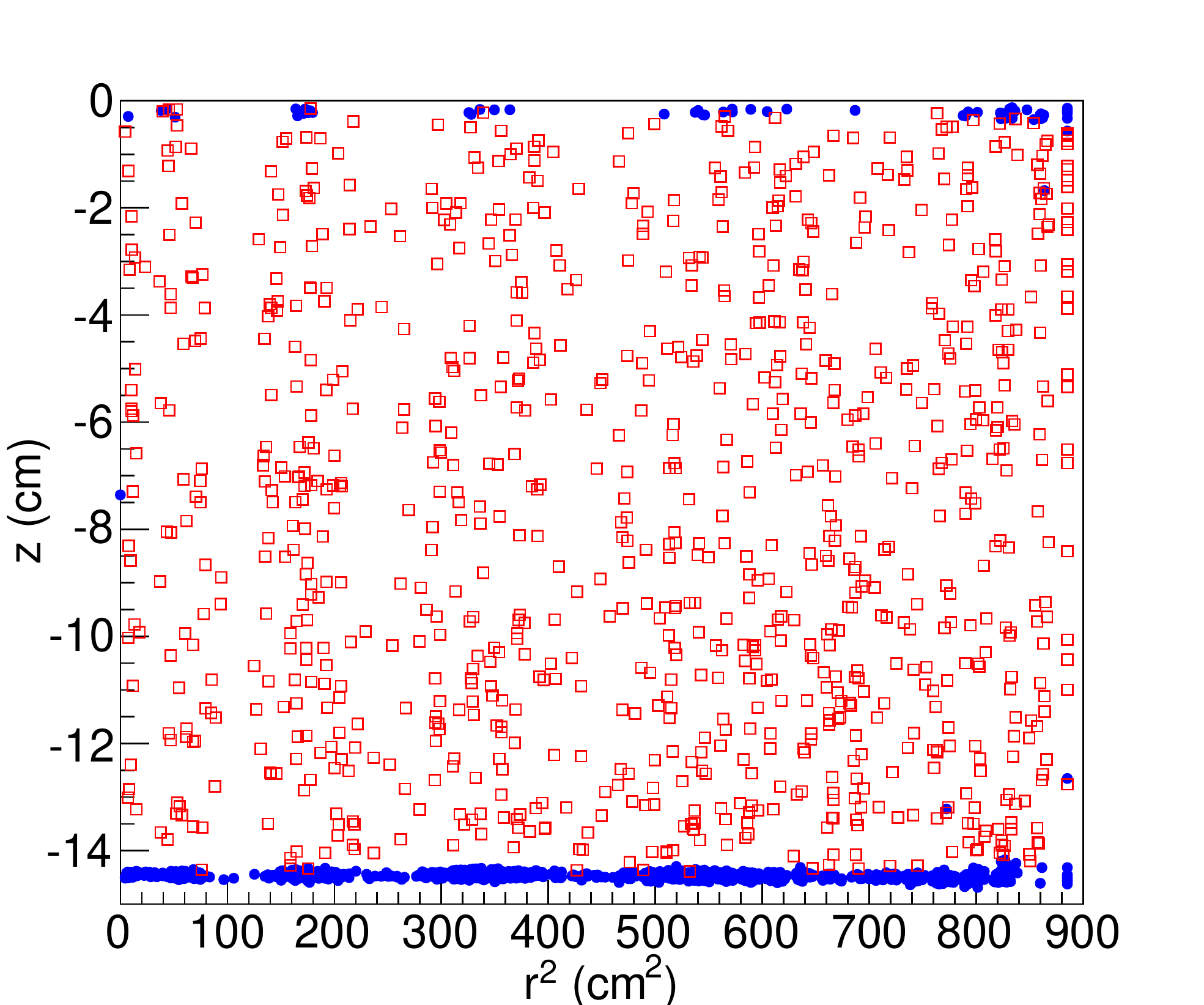}
}
\caption{Distributions of $^{214}$Bi-$^{214}$Po delayed coincidence events a) in delay
time, b) in $\Delta L^2$, and c) in $\alpha$ position for clusters A (blue dot) 
and B (red open square), cluster A distributed at the top and bottom mainly
while cluster B distributed uniformly in between.}
\label{fig:Rn222_cross_band}
\end{figure}
\begin{figure}[!htbp]
\centering
  \includegraphics[width=.7\linewidth]{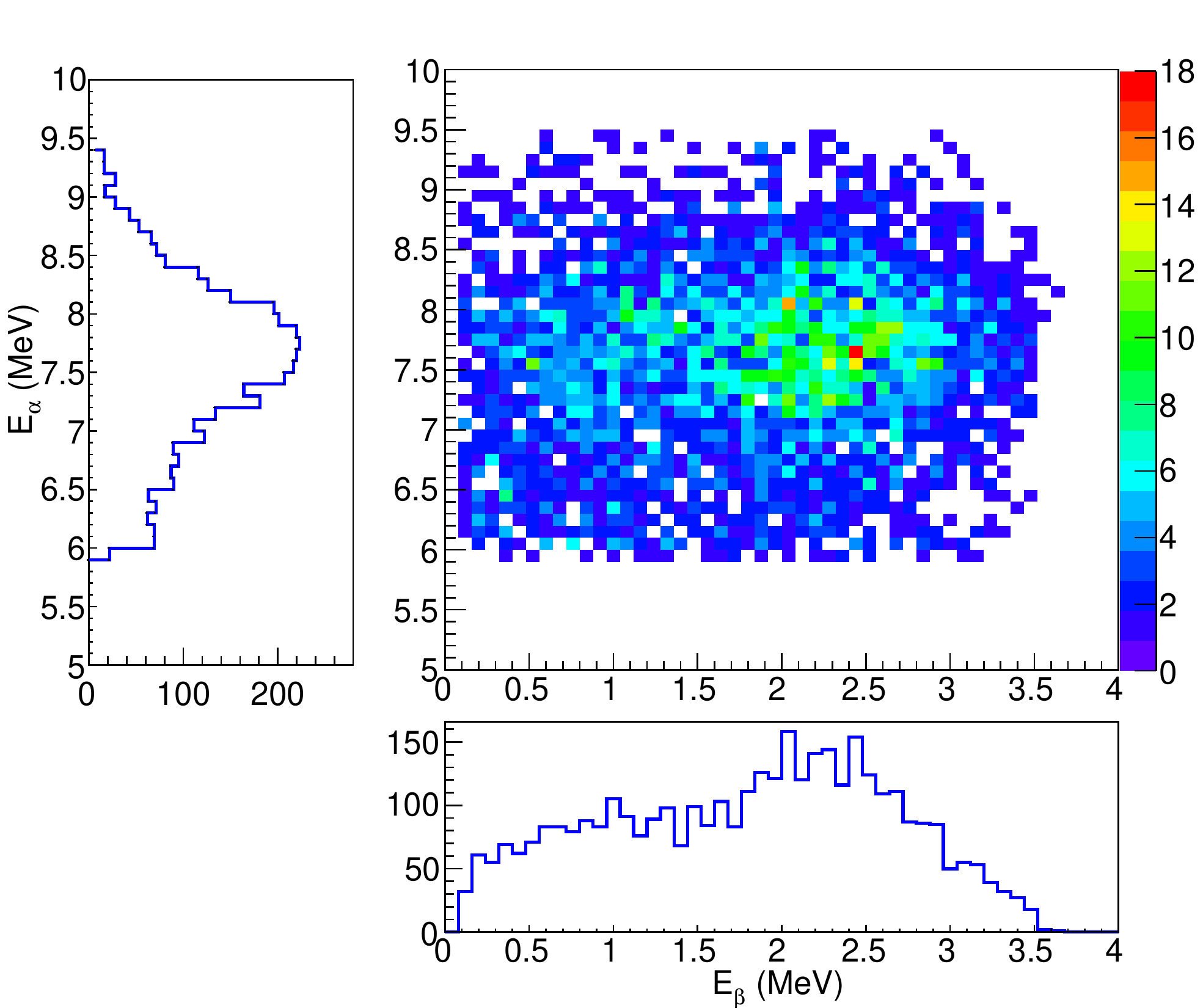}
\caption{Combined energy of $^{214}$Po $\alpha$ vs $\beta$ energy distribution. }
\label{fig:Bi214Po214_energy}
\end{figure}

With these selection cuts, we first searched for BiPo2E from the data.
Fig.~\ref{fig:Bi214Po214_band} shows the distributions of $\beta$ and $\alpha$
in the plane of log$_{10}$(S2/S1) vs. S1.
Two clusters in the $\alpha$ distribution can be clearly observed in Fig.~\ref{fig:Po214_band}, 
labeled as A and B in the figure.
The delay time and distance between $\beta$ and $\alpha$ in clusters A and B are shown
in Fig.~\ref{fig:Rn222_cross_band}, with 
event pairs in both clusters exhibit proper
timing and spatial correlations, therefore corresponding to genuine $\beta$-$\alpha$
coincidences. 
The position distribution of events in clusters A and B are shown
in Fig.~\ref{fig:Rn222_cross_Po214_position}.
Events in A are mostly located close to the cathode, while events in
B are uniformly distributed in the entire detector.
This phenomenon was first discussed in Ref.~\cite{EXO_Rn} for a liquid xenon TPC and 
explained by $^{214}$Bi$^+$ 
ions drifting toward and getting attached onto the cathode, a model which we 
shall refer to as the ``ion-drift model''.
Since the electric field near the cathode wires was much 
stronger than the average drift field and due to the short range of $\alpha$s, 
cluster A is displaced from cluster B in Fig.~\ref{fig:Po214_band}.
To reconstruct both clusters to the $^{214}$Po $\alpha$ energy, 
we derived an effective $\alpha$ energy reconstruction function as
\begin{equation}
E_{\alpha}(\rm{keV}) = \frac{\rm S1(PE)}{7.19 \rm{PE/keV}} + \frac{\rm S2(PE)}{116.88 \rm{PE/keV}}
\label{eq:comb_energy}
\end{equation}
The reconstructed $\beta$ and $\alpha$ energy distributions are shown in
Fig.~\ref{fig:Bi214Po214_energy}.
%
For BiPo1E event, we required that there was at least one S2 signal and
the energy cuts were the same with the above.
Similar as those in BiPo2E, $\alpha$ candidates were observed to form two clusters as in
Fig.~\ref{fig:Po214_band} with similar position distributions.
No time dependence was observed from BiPo1E and BiPo2E rates. 
Given that the underground radon level varied from 
tens to few hundred Bq/m$^{3}$ during the run, we concluded that the
$^{222}$Rn background was not due to an external air leak, but rather 
due to the internal surface emanation.

\begin{figure}[!htbp]
\centering
  \includegraphics[width=.7\linewidth]{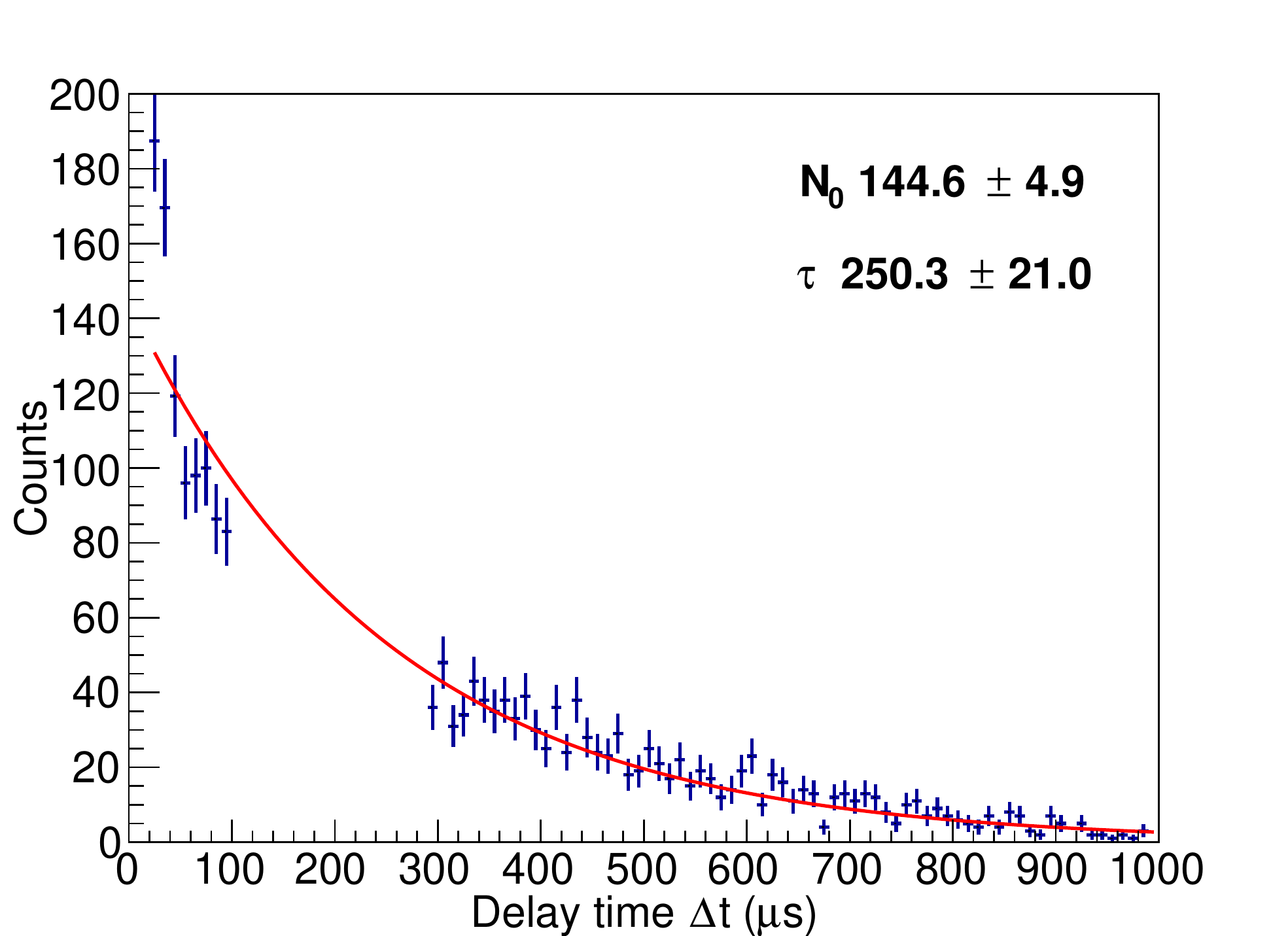}
\caption{Combined candidate events delay time distributions in the FV, where the gap 
between 100 to 300 $\mu$s is due to the delay time cuts for BiPo1E and BiPo2E. The uncertainty of the fitted lifetime includes both statistical and systematic uncertainties.}
\label{fig:Rn222_deltaT}
\end{figure}

Combining all the BiPo2E and BiPo1E candidate events, distributions of delay time
in the 54~kg fiducial volume (FV) is shown in Fig.~\ref{fig:Rn222_deltaT} 
and is fitted with a single exponential function. 
The data in the first two bins appear higher compared to the fit, 
implying a slight correlated background for small delay time but contributing 
only $\sim$3\% to the total delayed coincidence rate above the expected exponential curve.
The fitted decay time constant agreed well with the expectation.
The total radon rate can also be estimated by integrating 
the fit function.

The $^{222}$Rn decay rate based on BiPo1E, BiPo2E, or the combined fit methods
are summarized in Table~\ref{tab:Rn222_summary}.
\begin{table}[!htbp]
\centering
\caption{\label{tab:Rn222_summary}$^{222}$Rn level calculated from BiPo1E, BiPo2E and
combined fit in FV.}
\begin{tabular}[c]{r|ccc}
\hline\hline
Method                   &  BiPo1E   &    BiPo2E    &  Combined fit  \\
Delay time cut acceptance&   26.3\%  &    26.4\%    &  100.0\%   \\
$\beta$ energy cut       &   98.2\%  &    98.2\%    &   98.2\%     \\
$\alpha$ energy cut      &  100.0\%  &   100.0\%    &  100.0\%     \\
Branching ratio          &  99.98\%  &   99.98\%    &  99.98\%     \\
\hline
\multirow{2}{*}{$^{222}$Rn level}  & 0.66 mBq  & 0.79 mBq & 0.68$\pm$0.13 mBq \\
                                   & 12.3 $\mu$Bq/kg & 14.7 $\mu$Bq/kg & 12.5$\pm$2.4 $\mu$Bq/kg \\
\hline\hline
\end{tabular}
\end{table}
The signal selection efficiencies were estimated by the MC simulation. 
The delay time cut
acceptances for BiPo1E and BiPo2E were 26.3\% and 26.4\%, whereas that for
the combined fit is 100\% due to the integration range from zero to infinity.
The accidental background was estimated similarly as in Sec.~\ref{sec:Kr} and confirmed 
to be negligible. 
The mean $^{222}$Rn level was obtained using the combined fit, 
and the uncertainty was estimated based on the largest difference among three methods.
This result is at a similar level as in the XENON100 and LUX
experiments~\cite{XENON100_bk,XENON100_Rn,LUX_bk,LUX_Rn,LUX_Rn2}, and such an internal background
will pose challenge to next generation of liquid xenon experiment.

As mentioned earlier, $\beta$-decays in the $^{222}$Rn chain contribute to
the low energy background. 
To properly estimate the off-equilibrium $\beta$-decay contribution downstream of 
$^{210}$Pb, 
in the MC simulation, 
$^{222}$Rn events were assumed to be produced uniformly in position in the liquid xenon (so were all the progenies) and in a duration same as the entire period of the experiment, 
and were let decay all the way.
Events that passed all selection cuts and fell into the dark matter data taking period
were counted as background.
Based on the $^{222}$Rn level in the FV, the mean ER background contribution
to the PandaX-I experiment for each $\beta$-decay progeny
is summarized in Table \ref{tab:Rn222_mDRU}. 
\begin{table}[h!]
\centering
\caption{\label{tab:Rn222_mDRU}Background contribution from $^{222}$Rn.}
\begin{tabular}[c]{r|cr}
\hline\hline
Isotope & Background (mDRU) \\
\hline
$^{214}$Pb & 0.17  \\
$^{214}$Bi & 0.002  \\
$^{210}$Pb & 0.11  \\
$^{210}$Bi & 0.03  \\
$^{218}$Po & 0.002  \\
\hline
Total      & 0.32$\pm$0.06  \\
\hline\hline
\end{tabular}
\end{table}

\section{$^{220}$Rn}
\label{sec:Rn220}
\begin{figure}[!htbp]
\centering
  \includegraphics[width=.7\textwidth]{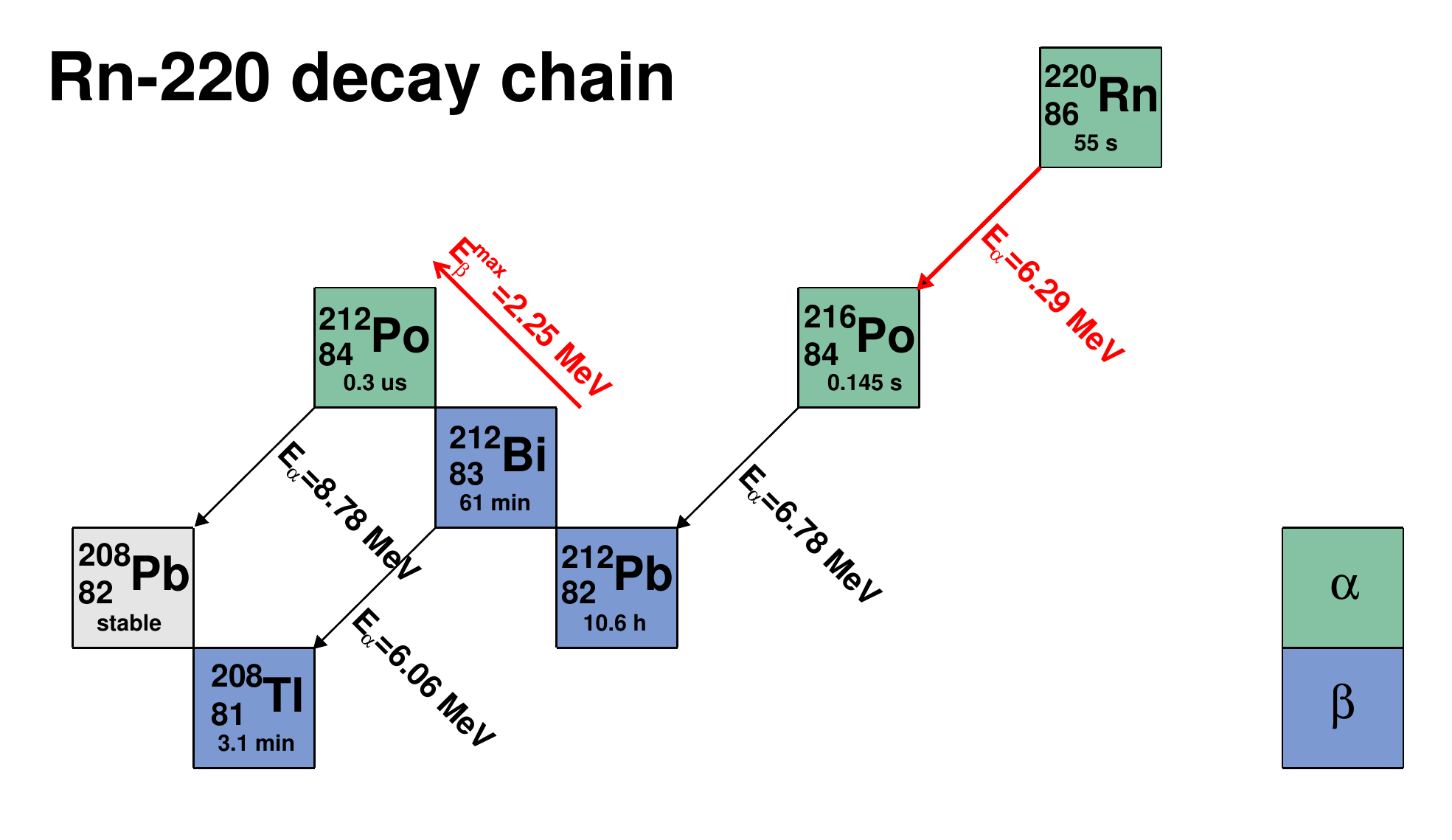}
  \caption{$^{220}$Rn decay chain. The red arrows indicate the $\beta$-$\alpha$ and $\alpha$-$\alpha$ delayed coincidences used 
    in this analysis.}
  \label{fig:rn220_decay_chain}
\end{figure}

$^{220}$Rn is the decay progeny of $^{232}$Th with a decay half-life of 55 s.
The decay chain of $^{220}$Rn is illustrated in Fig.~\ref{fig:rn220_decay_chain}.
There are two delayed coincidences which can be used as clean tags, 
$^{212}$Bi-$^{212}$Po $\beta$-$\alpha$ and  $^{220}$Rn-$^{216}$Po $\alpha$-$\alpha$
delayed coincidences.
$^{212}$Bi has a 64.06\% probability to $\beta$-decay, with a maximum energy 
of $\beta$ of 2.25 MeV. The daughter $^{212}$Po decays with a half-life of 
0.3 $\mu$s into a stable element $^{208}$Pb
emitting an $\alpha$ particle of 8.78~MeV.
The half-life of $^{216}$Po is 0.145 second and the $\alpha$ energy 
is 6.29 MeV and 6.78 MeV for $^{220}$Rn and $^{216}$Po, respectively.
The two $\alpha$ decays were recorded in different events.

%

\begin{figure}[!htbp]
  \centering
  \includegraphics[width=.5\linewidth]{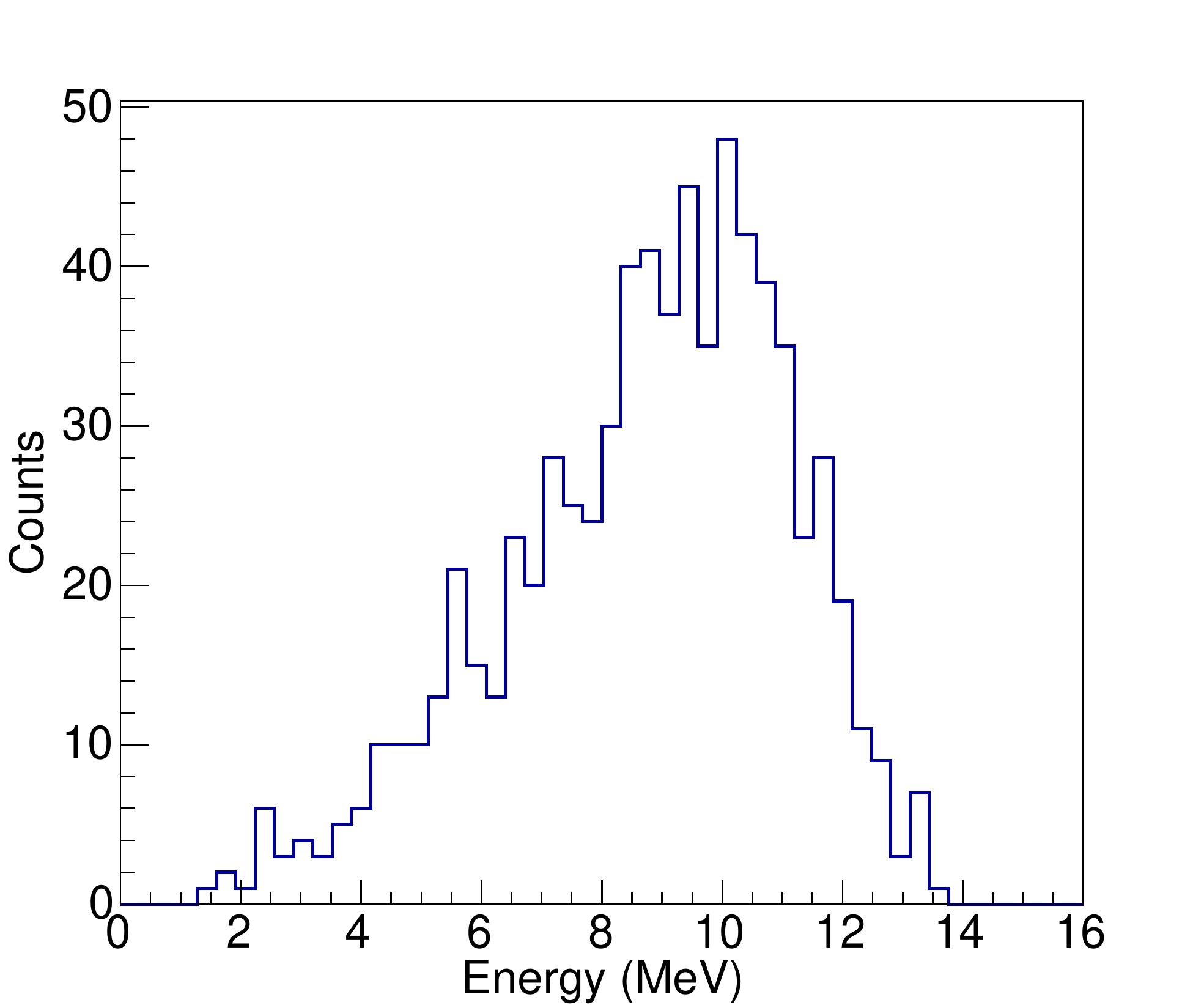}
  \caption{Distribution of the reconstructed energy for the $^{212}$Po $\alpha$s.}
  \label{fig:Po212_energy}
\end{figure}

Similar to the selection of BiPo1E events in $^{222}$Rn, the event selection
energy cuts for $^{212}$Bi-$^{212}$Po were set in the range from 
100 keV to 3 MeV for $\beta$s and 
$>$3 MeV for $\alpha$s, and the delay time cut was set between 0.3 to 1.0 $\mu$s. 
Similar to $^{214}$Po in Fig.~\ref{fig:Po214_band}, distribution of selected $^{212}$Po
$\alpha$ events also has two separated clusters, corresponding to the cathode
and bulk $\alpha$s.
The $\alpha$ energy was reconstructed using Eq.~\ref{eq:comb_energy}, displayed 
in Fig.~\ref{fig:Po212_energy}.
Due to the short decay of $^{212}$Po, the mean $\alpha$ energy 
is larger than the expected 8.78~MeV due to some pileup of S2s from the  
$\beta$ and $\alpha$.

\begin{figure}[!htbp]
\centering
\subfigure[$^{220}$Rn]
{
  \includegraphics[width=.45\linewidth]{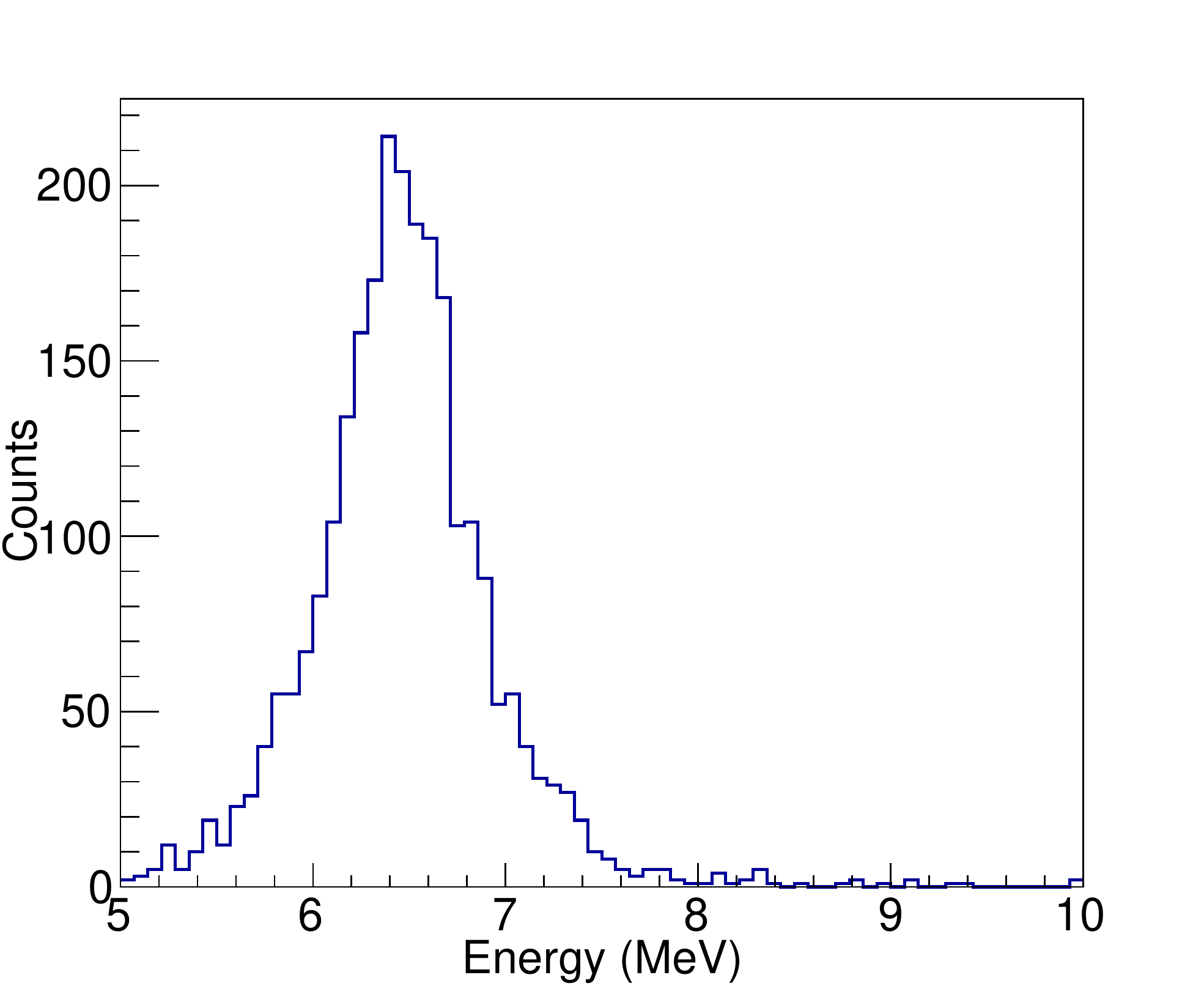}
}
\subfigure[$^{216}$Po]
{
  \includegraphics[width=.45\linewidth]{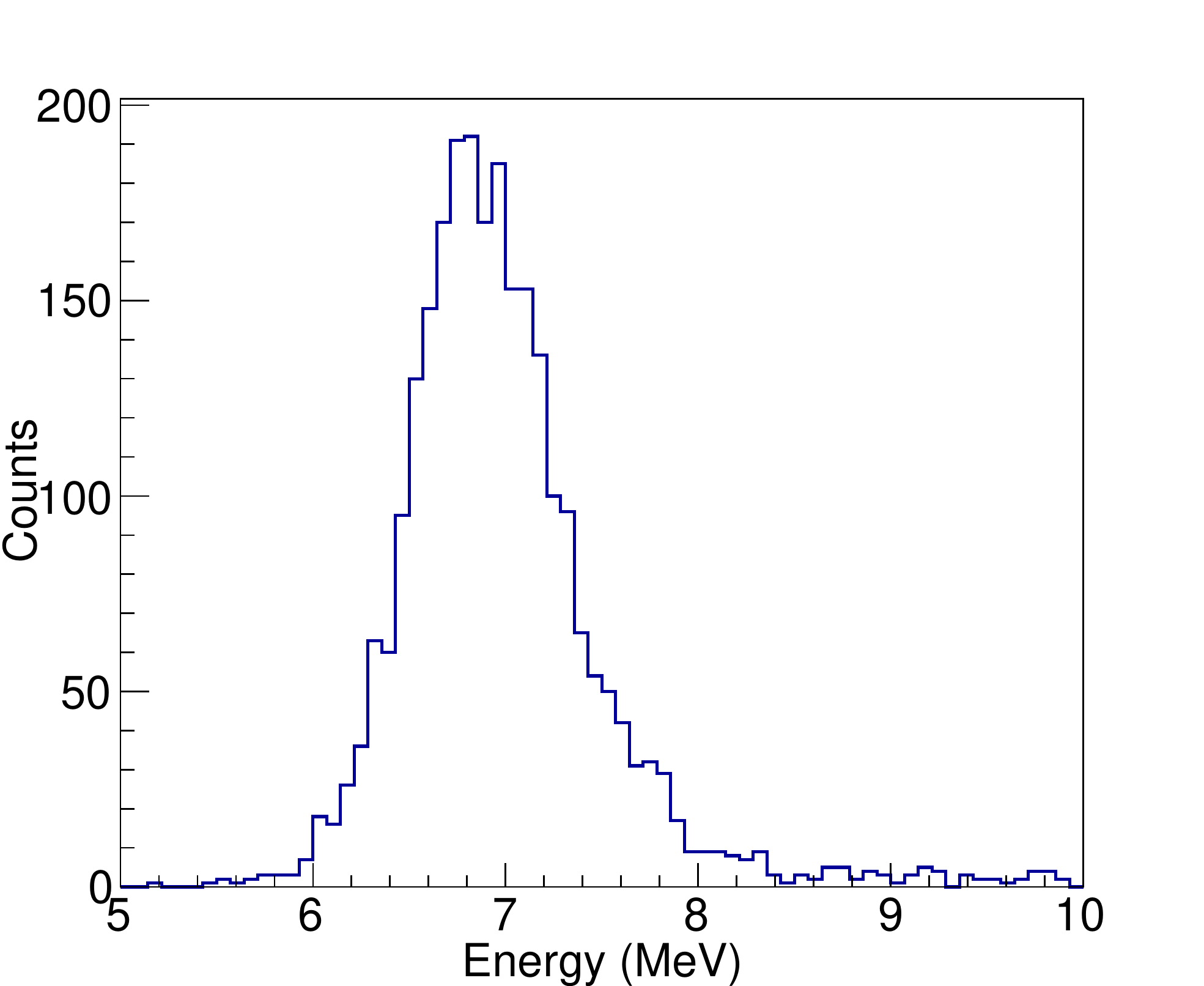}
}
\caption{Reconstructed energy of $\alpha$s from $^{220}$Rn (a) and $^{216}$Po (b).}
\label{fig:Rn220_RnPo_energy}
\end{figure}

\begin{figure}[!htbp]
\centering
\subfigure[Delay time]
{
  \includegraphics[width=.45\linewidth]{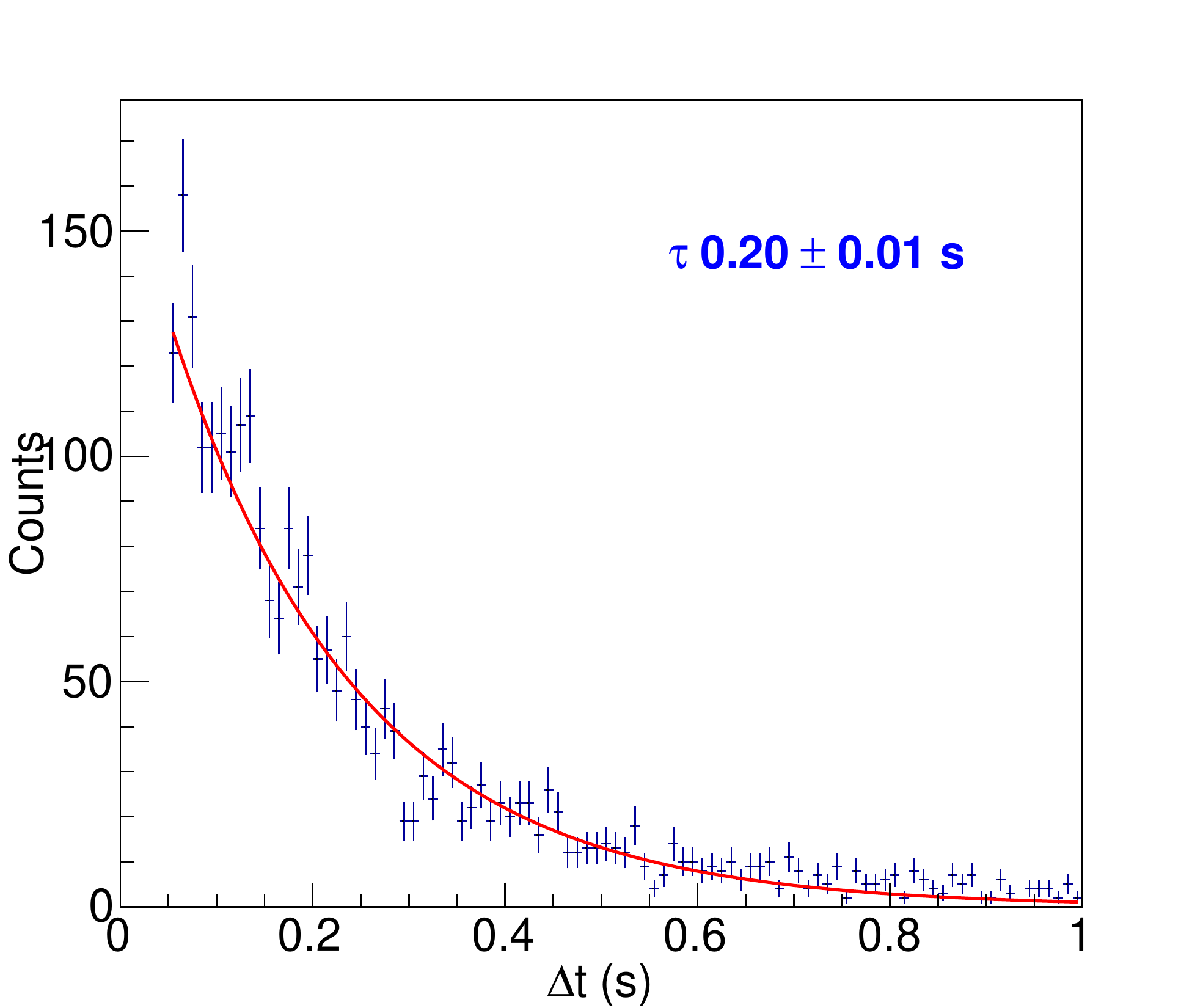}
}
\subfigure[$\Delta L^2$]
{
  \includegraphics[width=.45\linewidth]{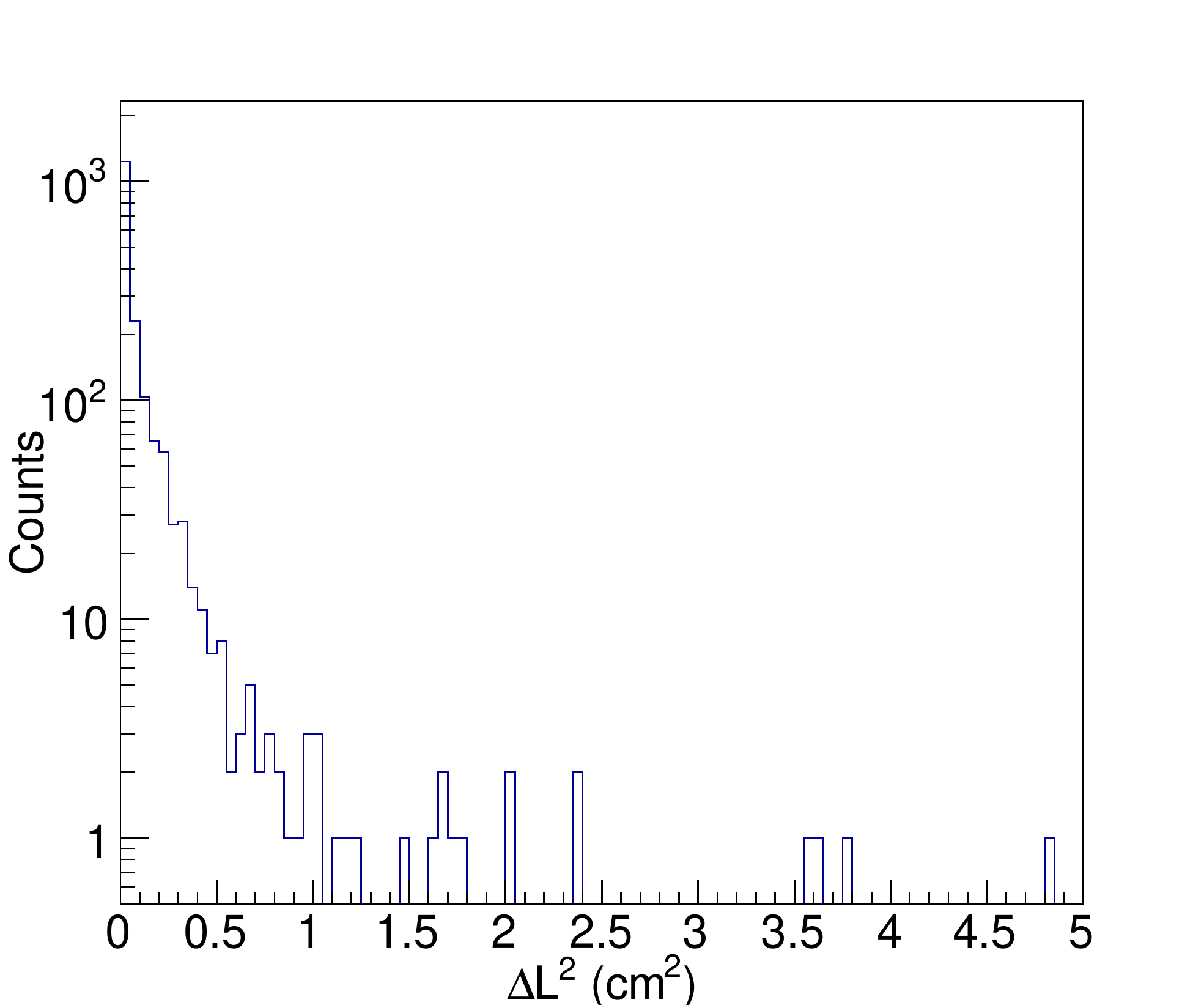}
}
\caption{$^{220}$Rn-$^{216}$Po delay time (a) and distance (b) distributions. The fitted
mean lifetime $\tau$ should be compared to the nominal value of 0.209 s.}
\label{fig:alpha_alpha}
\end{figure}

\begin{figure}[!htbp]
\centering
\subfigure[$^{220}$Rn]
{
  \includegraphics[width=.45\linewidth]{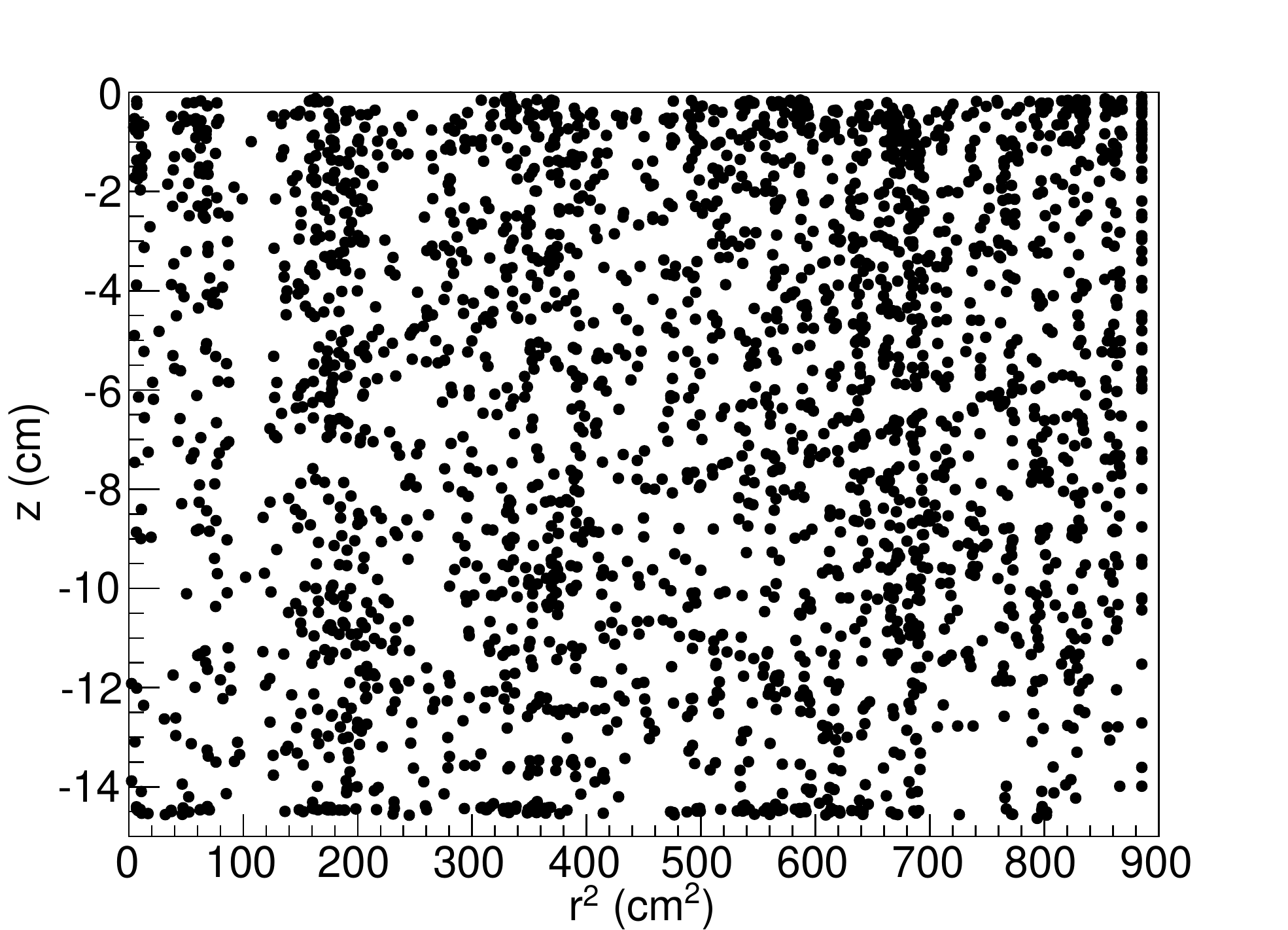}
}
\subfigure[$^{216}$Po]
{
  \includegraphics[width=.45\linewidth]{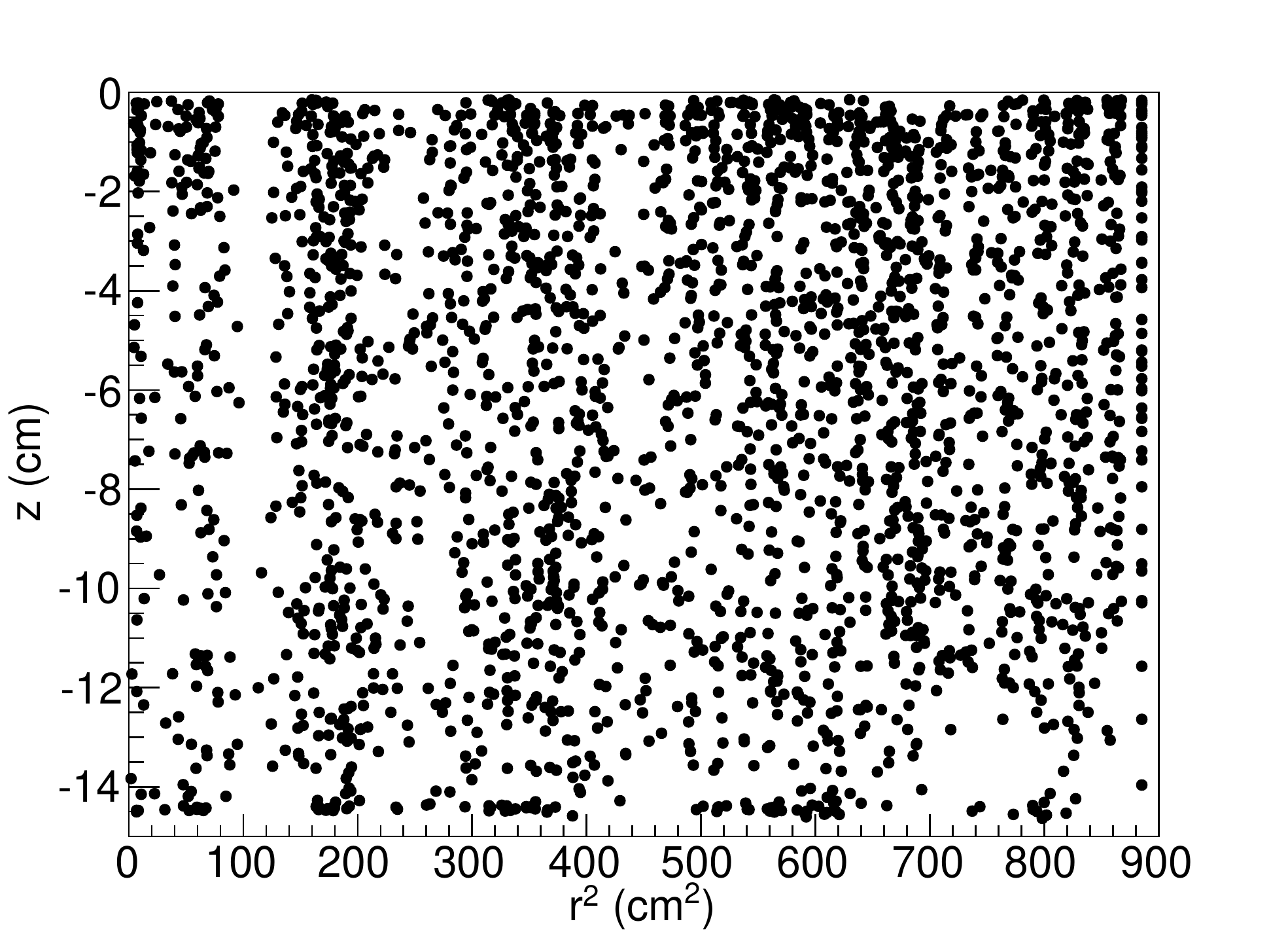}
}
\caption{Reconstructed position distributions of $^{220}$Rn-$^{216}$Po coincidence signals.}
\label{fig:Rn220Po216_spatial}
\end{figure}

To select the $^{220}$Rn-$^{216}$Po delayed coincidence, the $\alpha$ energy cuts were 
both
from 5~MeV to 12 MeV, and the delay time cut is from 0.05 to 1 s.
We required that each waveform should have only one S2 signal.
The distributions of energy, position and timing difference between the selected two $\alpha$s
are shown in Figs.~\ref{fig:Rn220_RnPo_energy} and \ref{fig:alpha_alpha}.
The delay time distribution agrees with the expectation and the position between
two $\alpha$s are close, consistent with the expectation too.
The position distributions of the two $\alpha$s in {\it z} and {\it r$^{2}$}
are shown in Fig.~\ref{fig:Rn220Po216_spatial}, where a much uniform distribution 
is found in comparison to the bismuth-polonium coincidences. 
This is again consistent with the ion-drift model in Ref.~\cite{EXO_Rn} that 
the early daughters are more uniformly distributed while the
later ones are more concentrated on the cathode since they have more time to drift
under the electric field. Like in $^{222}$Rn, 
no time dependence was observed in the $\beta$-$\alpha$ and 
$\alpha$-$\alpha$ rates, indicating that $^{220}$Rn was produced by internal surface
emanation. 

The $^{220}$Rn levels in the FV 
estimated from the two methods are summarized in 
Table \ref{tab:rn220_level}. 
\begin{table}[!htbp]
\centering
\caption{\label{tab:rn220_level}$^{220}$Rn level in FV.}
\begin{tabular}[c]{r|ccc}
\hline\hline
& BiPo $\beta$-$\alpha$ & RnPo $\alpha$-$\alpha$ \\
\hline
Delay time cut acceptance & 40.1\%            &  77.9\%    \\
$\beta$($\alpha$) energy cut& 95.7\%          & 100.0\%  \\ 
$\alpha$ energy cut   &      100.0\%          & 100.0\%  \\
Branching ratio       &      64.06\%          & 100\%    \\
\hline
$^{220}$Rn level      & 0.09 mBq     & 0.21 mBq     \\
$^{220}$Rn level      & 1.6 $\mu$Bq/kg  & 3.9 $\mu$Bq/kg  \\
\hline\hline
\end{tabular}
\end{table}
The accidental background was calculated to be negligible.
The two results 
differ by more than a factor of two, and that using the $\beta$-$\alpha$ method exhibits 
a strong depletion, consistent with the ion-drift model. 
The final level of $^{220}$Rn was estimated to be 2.7$\pm$1.1 $\mu$Bq/kg, in which
the average and half of the difference between the two methods were taken as the mean
and uncertainty, respectively. The statistical uncertainties from the two methods 
were both negligible.
Similar to Sec.~\ref{sec:Rn222}, 
$^{220}$Rn events and their daughters were placed uniformly in position 
in the liquid xenon in the simulation.
The contributions due to individual decay daughters to the low energy ER background
in the FV are listed in Table~\ref{tab:Rn220_mDRU}.
\begin{table}[h!]
\centering
\caption{Low energy background contribution from $^{220}$Rn daughters.}
\begin{tabular}[c]{r|cr}
\hline\hline
Isotope & Background (mDRU)  \\
\hline
$^{212}$Pb & 0.11  \\
$^{212}$Bi & 0.02  \\
$^{208}$Tl & 0.001  \\
\hline
Total      & 0.13$\pm$0.06  \\
\hline\hline
\end{tabular}
\label{tab:Rn220_mDRU}
\end{table}


\section{Single $\alpha$ studies}
\label{sec:SingleAlpha}
As shown in the coincidence analysis, $\alpha$s from the decay chain
can be easily distinguished due to their high energy and
large scintillation-to-ionization ratio. Therefore, single $\alpha$s 
(events in which only an isolated $\alpha$ was identified) can also
be used directly to estimate the radon background. In this approach,
we selected events with S1 larger than 2$\times$10$^{4}$ PE in the FV 
and required that the $z$ location encoded by the top/bottom ratio of the S1 signal 
was consistent with that obtained based on the timing difference 
between S1 and S2, to ensure that the S1 was paired with the correct S2 signal.

\begin{figure}[!htbp]
\centering
  \includegraphics[width=.7\linewidth]{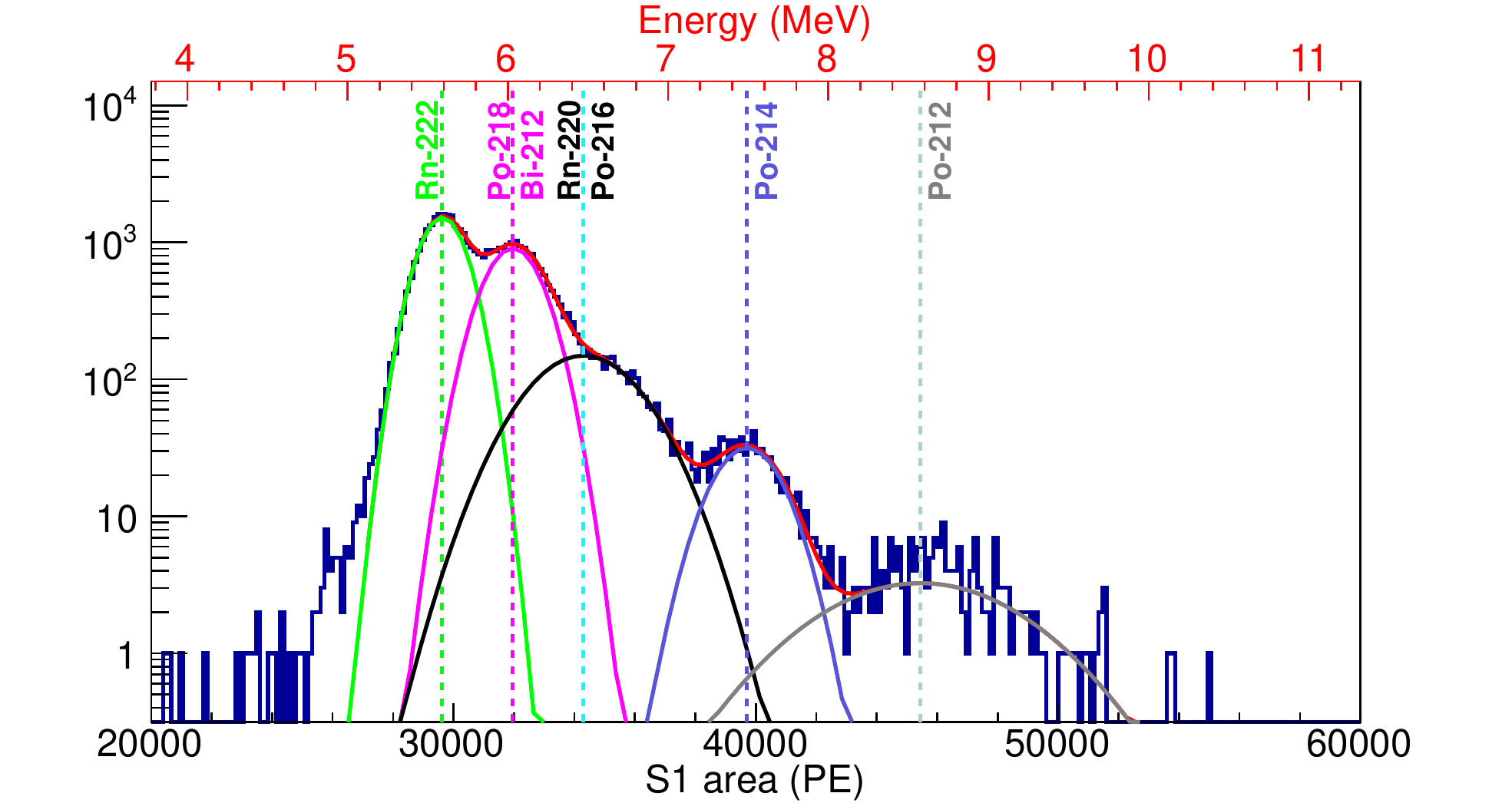}
\caption{Single $\alpha$ spectrum fitted with multi-Gaussian function. The vertical lines
indicate the peak positions. The upper x-axis in red is the reconstructed $\alpha$ energy.
}
\label{fig:single_alpha_spectrum_fit}
\end{figure}

Unlike in the coincidence measurement, identification of different single $\alpha$s relied on
good energy resolution. We used S1 signals corrected by an empirical position uniformity function 
to reconstruct $\alpha$ energy instead Eq.~\ref{eq:comb_energy}
to avoid saturation effects in S2s.
The S1 energy spectrum is shown in
Fig.~\ref{fig:single_alpha_spectrum_fit}.
The distribution is fitted with a sum of five Gaussian functions, with 
the peaks from low to high identified as $^{222}$Rn, $^{218}$Po (or $^{212}$Bi),
$^{220}$Rn (or $^{216}$Po), $^{214}$Po, and $^{212}$Po, with their centroids 
in reasonable agreement with expectation. Radon levels 
estimated from individual peaks are listed in Table~\ref{tab:single_alpha_summary}.
Along both chains, the event rate in the FV decrease towards later chain, consistent with the 
ion-drift model, also consistent with the findings in Refs.~\cite{EXO_Rn} 
and \cite{LUX_Rn,LUX_Rn2}.
We note that the values in Table~\ref{tab:single_alpha_summary} have sizeable discrepancies with
those in Tables~\ref{tab:Rn222_summary} and \ref{tab:rn220_level}. For example, 
$^{214}$Po (0.28 mBq) is different from $^{214}$Bi-$^{214}$Po coincidence (0.68 mBq),
and $^{212}$Po (0.06 mBq) is different from $^{212}$Bi-$^{212}$Po (0.09 mBq), indicating
systematic uncertainties in the single alpha analysis due to finite energy resolution 
to these highly energetic events. 
Therefore, in this paper, we report the values derived from the coincidence
method as the official results. Further studies are ongoing with the new data from PandaX-II in which
the TPC is a factor of four longer to investigate the difference between the two approaches.

\begin{table}[!htbp]
\centering
\caption{\label{tab:single_alpha_summary}Radon level from single $\alpha$ analysis with 
small statistical uncertainty ($<$10\%) for each $\alpha$ peak. The measured rate of the 6 MeV 
peak is more consistent with $^{218}$Po in the $^{222}$Rn chain so we did not attribute it to 
to $^{212}$Bi in the $^{220}$Rn chain.}
\begin{tabular}[c]{r|c|c|c|c}
\hline\hline
              Decay chain  &  Isotope   & $E_{\alpha, \rm{expected}}$(MeV) & $E_{\alpha, \rm{data}}$(MeV) & Radon rate (mBq) \\
\hline
\multirow{3}{*}{$^{222}$Rn} & $^{222}$Rn &  5.49  & 5.59 & 3.9 \\
\cline{2-5}                & $^{218}$Po &  6.00  & 6.02 & 2.8 \\
\cline{2-5}                & $^{214}$Po &  7.69  & 7.69 & 0.28 \\
\hline\hline
\multirow{3}{*}{$^{220}$Rn}& $^{220}$Rn &  6.29  & \multirow{2}{*}{6.47} & \multirow{2}{*}{0.9} \\
\cline{2-3}                & $^{216}$Po &  6.78  &      &  \\
\cline{2-5}                & $^{212}$Po &  8.78  & 8.58 & 0.06 \\
\hline\hline
\end{tabular}
\end{table}

\section{Summary}
\label{sec:sum}
We present a krypton and radon background study based on the PandaX-I 
full exposure data.
By searching for delayed coincidence events along their decay chains, 
we obtained the krypton and radon decay rates in the detector.
With an updated data selection and MC simulation, the expected background due to $^{85}$Kr,
$^{222}$Rn and $^{220}$Rn were estimated to be 2.0$\pm$0.6,
0.32$\pm$0.06 and 0.13$\pm$0.06 mDRU respectively, consistent with
and should supersede the initial estimates presented in Ref.~\cite{pandaxI2nd}.
These background rates appeared stable over time and distributed throughout the 
detector. For increasingly large detectors in the next generation experiments, 
such background may become more important, and special care is called for to control such background to a new level.

%
%
%
%

\section{Acknowledgement}
  This work has been supported by a 985-III grant from Shanghai
  Jiao Tong University, grants from National Science Foundation of
  China (Grant Nos. 11435008, 11455001, 11505112 and 11525522), and 
  grants from the Ministry of Science and Technology of 
  China (Grant Nos. 2016YFA0400301 and 2016YFA0400302).  This work 
  is supported in part by the Key Laboratory for Particle Physics, Astrophysics and
  Cosmology, Ministry of Education, Shanghai Key Laboratory for Particle
  Physics and Cosmology (SKLPPC), and the 
  Chinese Academy of Sciences Center for Excellence in Particle Physics (CCEPP).  







\end{document}